\documentclass[preprint,12pt,3p]{elsarticle}

\usepackage{amssymb}

\usepackage{amsmath}
\usepackage{tcolorbox}
\usepackage{xspace}
\usepackage{enumitem}

\usepackage{amsfonts}
\usepackage{url}

\usepackage{multirow}
\usepackage{tabularx}
\newcolumntype{C}[1]{>{\centering\arraybackslash}p{#1}}

\usepackage{graphicx}  %插入图片的宏包
\usepackage{float}  %设置图片浮动位置的宏包
\usepackage{subfigure}  %插入多图时用子图显示的宏包

\usepackage{booktabs} % 导入三线表需要的宏包
\usepackage{multirow}
\usepackage{adjustbox}

\newcommand{\tool}{\textsc{TigAug}\xspace}
\newcommand{\todo}[1]{\textcolor{black}{#1}}
\newcommand{\ly}[1]{\textcolor{black}{#1}}

\journal{Journal of Systems and Software}

\begin{document}

\begin{frontmatter}



\title{\tool: Data Augmentation for Testing Traffic Light Detection in Autonomous Driving Systems}

\author[1]{You Lu\fnref{fn1}}
\ead{ylu24@m.fudan.edu.cn}

\author[1]{Dingji Wang\fnref{fn1}}
\ead{wangdj25@m.fudan.edu.cn}

\author[2]{Kaifeng Huang} 
\ead{kaifengh@tongji.edu.cn}

\author[1]{Bihuan Chen\fnref{fn1}\corref{cor1}}
\ead{bhchen@fudan.edu.cn}

\author[1]{Xin Peng\fnref{fn1}}
\ead{pengxin@fudan.edu.cn}

\affiliation[1]{organization={School of Computer Science, Fudan University},%Department and Organization
            % addressline={a}, 
            city={Shanghai},
            country={China}}

\affiliation[2]{organization={School of Software Engineering, Tongji University},%Department and Organization
            % addressline={a}, 
            city={Shanghai},
            country={China}}

\fntext[fn1]{Y. Lu, D. Wang, B. Chen and X. Peng are also with the Shanghai Key Laboratory of Data Science.}
% \affiliation[3]{organization={Shanghai Key Laboratory of Data Science},%Department and Organization
%             % addressline={a}, 
%             city={Shanghai},
%             % postcode={}, 
%             % state={},
%             country={China}}               
\cortext[cor1]{Corresponding author}

%% Abstract
\begin{abstract}
    % !TeX root = ../main.tex
Autonomous vehicle technology has been developed in~the last decades~with recent advances~in sensing and computing technology. There is an urgent need to ensure the~reliability~and robustness of autonomous driving systems (ADSs). Despite the recent achievements in testing various ADS modules, little attention has been paid on the automated testing~of~traffic light detection models in ADSs. A common practice~is~to~manually~collect and label traffic light data. However, it is labor-intensive, and even impossible to collect diverse data under different driving environments.

To address these problems, we propose and implement~\tool to automatically augment labeled traffic light images for testing traffic light detection models in ADSs. We construct two~families~of~metamorphic relations and three families of transformations based~on~a systematic understanding of weather environments, camera properties, and traffic light properties. We use augmented images~to~detect erroneous behaviors of~traffic~light~detection~models by transformation-specific metamorphic~relations,~and~to~improve the performance~of~traffic light detection models~by~retraining. Large-scale experiments~with four state-of-the-art traffic light detection models and two traffic light datasets have demonstrated that i) \tool is effective~in~testing traffic light detection models, ii) \tool is efficient~in synthesizing traffic light images,~and iii) \tool~generates traffic light images with acceptable naturalness.

\end{abstract}

\begin{graphicalabstract}
\includegraphics[width=\textwidth]{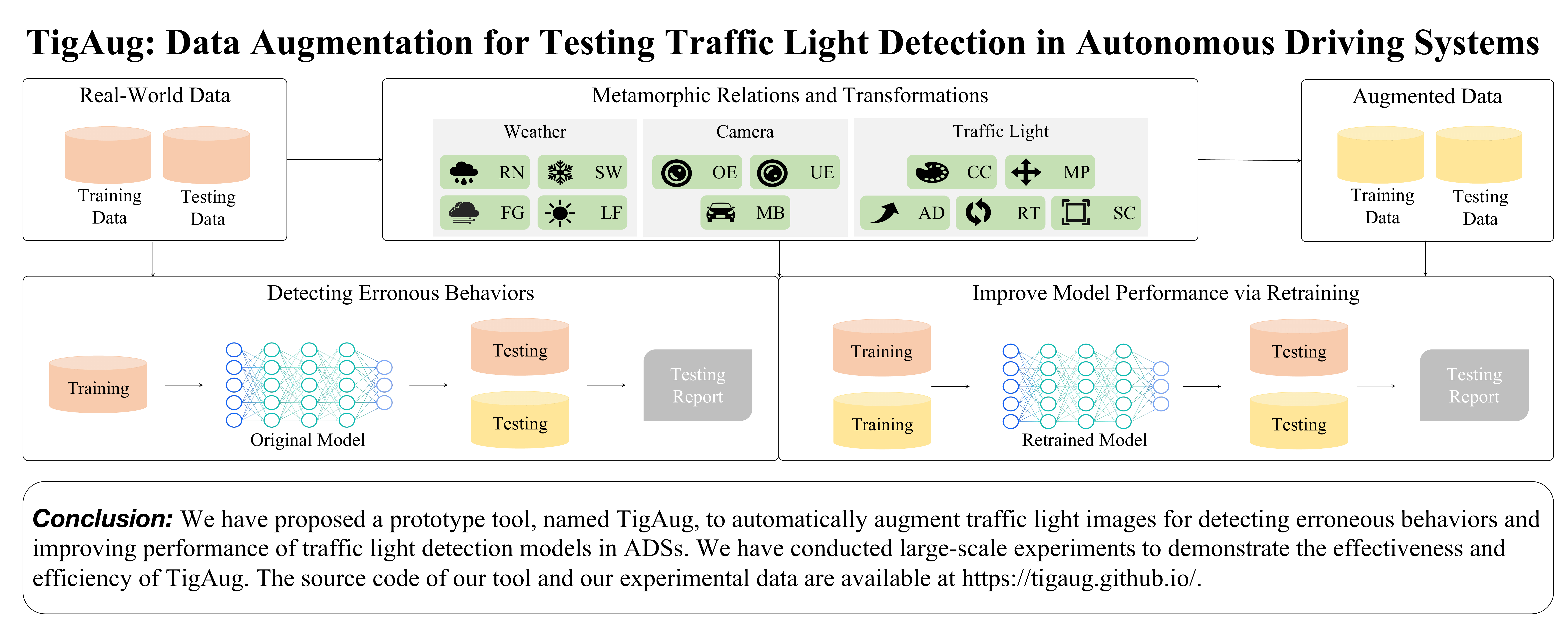}
\end{graphicalabstract}

\begin{highlights}
  \item We constructed two families of metamorphic relations~and three families of transformations.

  \item Implementation of a prototype~tool, named \tool, to automatically augment traffic light~images.
    
  \item Demonstration of effectiveness and efficiency of \tool.
\end{highlights}

%% Keywords
\begin{keyword}
Mutation Testing \sep Testing of Deep-Learning Models
\end{keyword}

\end{frontmatter}



% !TeX root = ../main.tex

\section{Introduction}

Academic and industrial efforts have been increasingly devoted~to developing autonomous vehicle technology~in~the last decades. These developments have been fueled by recent advances~in sensing and computing technology together~with the~impact~on~automotive transportation and the benefit~to~society (e.g., reducing~vehicle collisions, and providing personal mobility to disabled people)~\cite{paden2016survey}.~The~automation~system~of autonomous vehicles, also known as autonomous driving~system (ADS), is typically organized into two main parts,~i.e.,~the perception system and the decision-making~system~\cite{badue2021self}.~The~perception~system estimates the vehicle and environment state~using the data captured by on-board sensors such as camera~and LIDAR,~while the decision-making system navigates the vehicle from its initial position to the final destination. %specified by the user. 

As a safety-critical system, it is important to ensure~the~reliability and robustness of an ADS. Unfortunately, the state-of-the-practice ADSs from leading~companies such as Tesla,~Waymo and Uber~are still vulnerable to corner cases and exhibit~incorrect behaviors,~due to the extremely complicated~and~diverse real-world driving environments. These incorrect behaviors~may lead~to~catastrophic~consequences and unsustainable losses, as~evidenced~by~many~reported~traffic~accidents~\cite{tesla, waymo, uber}. Therefore, on-road testing is adopted~by these leading companies to achieve quality assurance for ADSs. To further test extreme conditions that are difficult~or expensive to produce in real-world environments, simulation testing is also widely adopted by these~leading~companies~\cite{kaur2021survey,ji2021perspective}.

In recent years, many testing approaches have been developed~by the software engineering community to ensure~the~reliability~and~robustness of ADSs. Specifically, one~main~line~of work attempts~to~apply search-based testing to detect safety~violations for ADSs~\cite{ben2016testing, abdessalem2018testinga, abdessalem2018testingb, gladisch2019experience, han2020metamorphic, li2020av, tian2022mosat, sun2022lawbreaker, gambi2019automatically, zhong2022detecting}.~They~formulate a test scenario~as~a~vector of variables (e.g., vehicle speed and fog degree), and apply~generic~algorithms~to~search for test~scenarios that violate safety requirements. Another main thread of work focuses on testing DNN (deep neural network)-based modules in ADSs. They use metamorphic~testing to generate images of driving scenes \cite{pei2017deepxplore, tian2018deeptest, shao2021testing, wang2020metamorphic} and point clouds~of driving scenes \cite{guo2022lirtest}. For example,~DeepTest~\cite{tian2018deeptest}~uses~weather and affine transformations to synthesize images for testing~steering angle decision models in ADSs. Similarly, LiRTest~\cite{guo2022lirtest}~applies weather and affine transformations~to~synthesize point clouds for testing 3D object detection models in ADSs.

Despite these advances in finding erroneous behaviors in various modules of ADSs, little attention has been paid~on~the testing~of~traffic light detection models in ADSs. Traffic~lights are~used to control the movement of traffic, and thus play~a vital role in ensuring traffic safety. Therefore, ADSs employ traffic light detection~models (e.g., \textit{YOLO}~\cite{redmon2017yolo9000}, \textit{Faster R-CNN} \cite{ren2015faster} and \textit{SSD}~\cite{liu2016ssd})~to~detect~the~position~of~one~or~more~traffic lights in the driving scene (e.g., represented in an image)~and recognize their states (e.g., red, green,~and yellow)~\cite{badue2021self}.~When an ADS fails to correctly recognize traffic lights, it may~cause serious traffic accidents. For example,~Tesla's~Autopilot~misidentified the moon as a yellow light, causing the vehicle~to unexpectedly slow down at highway speeds~\cite{tesla-light}. Uber's~autonomous vehicle passed through a red light three seconds~after~the light had turned red and while a pedestrian was in the crosswalk~\cite{uber-light}. Therefore, it is crucial to test traffic light detection in ADSs.

The testing of traffic light detection heavily relies on the labeled traffic light data (i.e., images of traffic lights), which~is usually manually collected. Specifically,~on-road~testing~is~employed to capture images of traffic lights via cameras. However, it is resource-intensive or even impossible to collect diverse data under different driving~environments. Then, the captured traffic light images~are~manually labeled to mark the position and state of traffic lights. However, it is a labor-intensive and time-consuming task, especially when the number of traffic light images increases. To the best of our knowledge, Bai et al.'s work~\cite{bai2021metamorphic} is the first and only work to automatically generate synthetic images of traffic lights. However,~they~only~consider one transformation scenario, i.e., changing the color of traffic lights, hindering the diversity of generated traffic light images. % and hurting the effectiveness of traffic light detection testing.

To address these problems, we propose and implement~a systematic data augmentation approach, named \tool,~to~automatically augment labeled traffic light images for testing traffic light detection models in ADSs. We~systematically build~two families of metamorphic relations~and~three~families of transformations based~on~a systematic understanding~of weather environments,~camera properties, and traffic~light~properties. Given~a labeled traffic light image from real world, \tool applies transformations~to~synthesize~augmented~traffic light images. Then, it uses transformation-specific metamorphic relations between the real-world image~and~those~augmented images to identify erroneous behaviors of traffic light detection models in ADSs. Moreover, the augmented images can be used to retrain and improve traffic light detection models. 

We conduct large-scale experiments to evaluate \tool, using four state-of-the-art traffic light detection models and two~traffic light datasets. First, we apply transformations~on~the testing data, and the mean average precision~of~the~original~models suffers~a~decrease of \todo{39.8\%} on the augmented~testing~data. Second,~we~obtain~retrained models by adding~augmented~training data, and~the~mean average precision of the~retrained~models  achieves~an~increase~of~\todo{67.5\%} on the augmented testing~data. These results demonstrate that \tool is effective~in~detecting erroneous behaviors and improving the performance of traffic light detection models in ADSs.~Third,~we measure the time overhead of data augmentation and retraining~to~evaluate~the efficiency of \tool, and it takes on average \todo{0.88} seconds to synthesize an image and \todo{36} hours to retrain a model,~which~is acceptable given the improved mean average~precision.~Finally, we manually identify unnatural images from the synthesized images, and \todo{27.9\%} of the synthesized images from~each~transformation are considered as unnatural, but they~have~little~impact on model performance. It indicates that our synthesized traffic light images can be directly used without manual cleaning.

In summary, this work makes the following contributions.
\begin{itemize}[leftmargin=*]
\item We constructed two families of metamorphic relations~and three families of transformations to simulate the impact of weather~environments, camera properties and traffic light properties on the captured real-world traffic light images.

\item We implemented the proposed approach as a prototype~tool, named \tool, to automatically augment traffic light~images for detecting erroneous behaviors and improving the performance~of~traffic light detection models in ADSs.

\item We conducted large-scale experiments, using four state-of-the-art traffic light detection models and two traffic light datasets, to show~the effectiveness and efficiency of \tool.
\end{itemize}

% !TeX root = ../main.tex

\section{Methodology}\label{sec:approach}

\begin{figure}[!t]
    \centering
    \includegraphics[width=\linewidth]{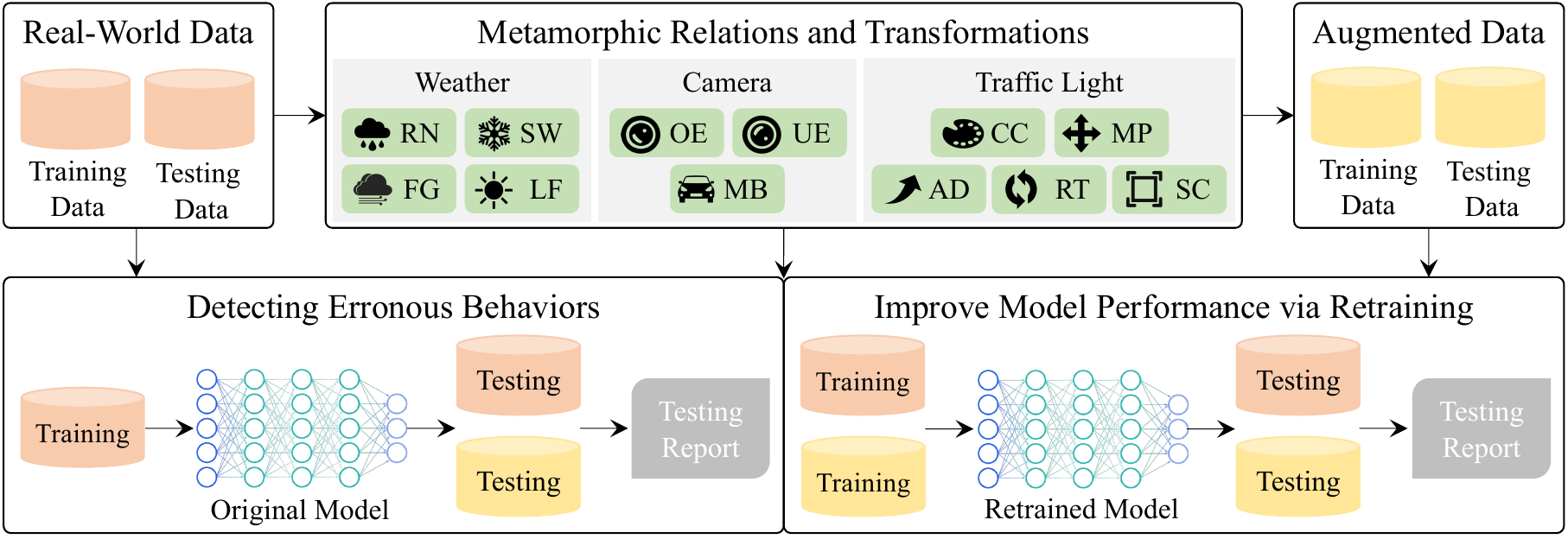}
    % \vspace{-5pt}
    \caption{Approach Overview of \tool}
%    \vspace{-5pt}
    \label{fig:overview}
\end{figure}

We design \tool as a systematic data~augmentation approach to automatically augment labeled traffic light images for testing traffic light~detection models in ADSs. Fig.~\ref{fig:overview} shows~an~overview of \tool. It is built~upon~our~domain understanding of how weather environments, camera~properties,~and~traffic light properties affect the traffic light images captured by cameras in real-world environments. We construct two families of metamorphic~relations (Sec.~\ref{sec:mr}) and three families of transformations (Sec.~\ref{sec:op}) with respect to weather environments, camera properties,~and~traffic light properties. 

On the basis of our domain analysis, \tool first applies transformations on labeled traffic light data from real world (i.e., real-world data) to generate augmented traffic light data (i.e., augmented data). Then, \tool uses transformation-specific metamorphic relations between the real-world testing data and those augmented testing data to identify erroneous behaviors of traffic light detection models trained from real-world training data. Moreover, \tool retrains traffic light detection models by adding augmented training data to real-world training data so as to improve model performance.

\subsection{Metamorphic Relations}\label{sec:mr}

A key challenge of traffic light data augmentation~is~to~automatically determine the expected output of traffic light~detection models on the augmented data. Metamorphic relations~\cite{chen2018metamorphic, segura2016survey} are known~to~be~capable of alleviating this test oracle challenge. In our scenario,~each metamorphic relation describes a property between~the~outputs of traffic light detection models on the~real-world traffic light data and the augmented~one. 

Specifically, for the three families~of transformations (Sec. \ref{sec:op}), we build two families of metamorphic relations. Before elaborating the metamorphic relations, we first formulate~relevant notations. For a traffic light~image $i$ from real-world data $\mathbb{I}$, the output of a traffic light detection model $m$~on~$i$ is denoted as $m[\![i]\!]$; and the augmented image after applying~a~transformation $\tau$ from weather transformations $\mathbb{W}$, camera transformations $\mathbb{C}$ or traffic light transformations $\mathbb{L}$ is denoted as $\tau(i)$.

Then, we formulate the metamorphic relation for weather and camera transformations by Eq.~\ref{eq:mr1},
\setlength{\abovedisplayskip}{2pt}
\setlength{\belowdisplayskip}{2pt}
\begin{equation}
%\begin{small}
\begin{aligned}\label{eq:mr1}
\forall~i \in \mathbb{I} \wedge \forall~\tau \in \mathbb{W} \cup \mathbb{C},~\mathcal{E}\{m[\![i]\!], m[\![\tau(i)]\!]\}
\end{aligned}
%\end{small}
\end{equation}
where $\mathcal{E}$ is a criterion asserting the equality of detection~outputs~on real-world image and augmented image. This metamorphic relation indicates that no matter how weather conditions and camera~effects are synthesized into a real-world image by applying weather~and camera transformations, the detection output of a traffic light detection model on the augmented~image $\tau(i)$ is expected to be consistent with that on the corresponding real-world image $i$. For example,~if the snow effect or the overexposure effect is synthesized into~a~real-world image $i$ to generate an augmented image $i^\prime$, the position and state of traffic lights in $i$ are the same to those in $i^\prime$, and hence the detection output on $i$ and $i^\prime$ should be identical. Otherwise,~erroneous behaviors are revealed by violating this metamorphic relation.

Similarly, we formulate the metamorphic relation for traffic light transformations by Eq.~\ref{eq:mr2}.
\setlength{\abovedisplayskip}{2pt}
\setlength{\belowdisplayskip}{2pt}
\begin{equation}
%\begin{small}
\begin{aligned}\label{eq:mr2}
\forall~i \in \mathbb{I} \wedge \forall~\tau \in \mathbb{L},~\mathcal{E}\{\tau(m[\![i]\!]), m[\![\tau(i)]\!]\}
\end{aligned}
%\end{small}
\end{equation}
This metamorphic relation indicates that as the position~and state~of traffic lights are changed by applying traffic light~transformations,~the detection output of a traffic light detection model on the augmented image $\tau(i)$ is expected~to~be~correspondingly changed by applying the transformation $\tau$ to the detection output on the corresponding real-world image $i$. For example, if a traffic light~in~a~real-world image $i$ is~changed from red to green to synthesize an augmented image $i^\prime$, the detection output on $i^\prime$ should be identical to that applying the same color change to the detection output on $i$.

Finally, we use mean average precision (mAP) to derive~the~equality criterion $\mathcal{E}$ because mAP is commonly used to evaluate object~detection systems, and it can compensate small drift of detected~bounding boxes from the ground truth bounding box. Specifically, mAP~is the mean of average precision (AP) scores of different object classes (e.g.,~different traffic light states in our scenario). AP is computed~by the~area under the precision-recall curve~\cite{everingham2009pascal}. Therefore, AP takes into account both precision and recall. To compute precision and recall, intersection over union (IoU) is used to measure the overlap between the detected bounding box and the ground truth bounding box. If IoU is greater than a threshold $\theta$ (e.g., 0.5), the detection is classified as a true positive; otherwise, the detection is classified~as~a false positive.   To take into account the impact of $\theta$, mAP@[.50,.95] is used to compute the average of mAP under 10 IoU thresholds~from 0.50 to 0.95 with a step size of 0.05~\cite{lin2014microsoft}. Hereafter, we use mAP to simplify mAP@[.50,.95] for the ease of presentation.

\subsection{Transformations}\label{sec:op}

To obtain more traffic light data exploring the input-output spaces of traffic light detection models with less labor and time cost, we implement twelve transformations to generate synthesized data that is close to real-world data in a small amount of time. As shown in Fig.~\ref{fig:overview}, these transformations are rain (RN), snow (SW),~fog~(FG), lens flare (LF), overexposure (OE), underexposure~(UE),~motion~blur (MB), changing color of traffic lights (CC), moving position of traffic lights (MP), adding traffic lights (AD), rotating traffic lights (RT),~and scaling traffic lights (SC). These transformations are classified into three families, i.e., weather transformations (to mimic~the~effects~of different weather environments), camera transformations (to simulate different camera effects), and traffic light transformations (to enrich different positions and states of traffic lights).

Before we clarify the detailed design~of~each transformation in each family, we first formulate relevant notations.~For~a real-world traffic light image, it contains a set of traffic~lights $T = \{t_1, t_2, ..., t_n\}$ with corresponding labels $L = \{l_1, l_2, ..., l_n\}$. The label $l$ of a traffic light $t$ is denoted as a 5-tuple $\langle x_{1}, y_{1}, x_{2}, y_{2}, state\rangle$, where $x_{1}$, $y_{1}$, $x_{2}$ and $y_{2}$ are used to represent the bounding box of the traffic light (i.e., a rectangular region around the traffic light~within the image), and $state$ represents the state of the traffic light (e.g.,~a stop state for a red traffic light).  $x_1$ and $y_1$ are the x and~y~coordinate of the top left corner of the bounding box, and $x_2$ and $y_2$ are the x and y coordinate of the bottom right corner of the bounding~box.

%\begin{center}
%    $Bounding\ Box = (xmin, ymin, xmax, ymax)$ .
%\end{center}

%\begin{center}
%    $xmin = top\_left\_x$,\\
%    $ymin = top\_left\_y$,\\
%    $xmax = xmin + width\_of\_bounding\_box$,\\
%    $ymax = ymin + height\_of\_bounding\_box$.\\
%\end{center}

% Some recent research illustrate that Deep Learning models can be tested and further fooled by adversarial perturbations. By adjusting various parameters, a perturbation $\delta$ can be added to the input $x_{i}$ such that the perturbed output $x_{i}' = x_{i} + \delta $ can be generated. But it's worth recalling that some perturbations may never happen in physical world (i.e., adding an abrupt square to the images). 

% In the light of real-world road traffic light images, we have to choose reasonable perturbations making the augmented images looks as real as possible. Meanwhile, the recent DeepTest and DeepRoad research use OpenCV and Generative Adversarial Networks (GANs) to generate road images with different weathers. These perturbations such as rain, snow and fog are simulating a real scenario which can be understood by human eyes, though the driving scenes generated by DeepTest and DeepRoad are still far from real-world scenes. 

\begin{figure*}[!t]
	\centering  
	% \subfigbottomskip=-3pt 
	% \subfigcapskip=-3pt 
	\subfigure[OI]{
		\includegraphics[width=0.11\linewidth]{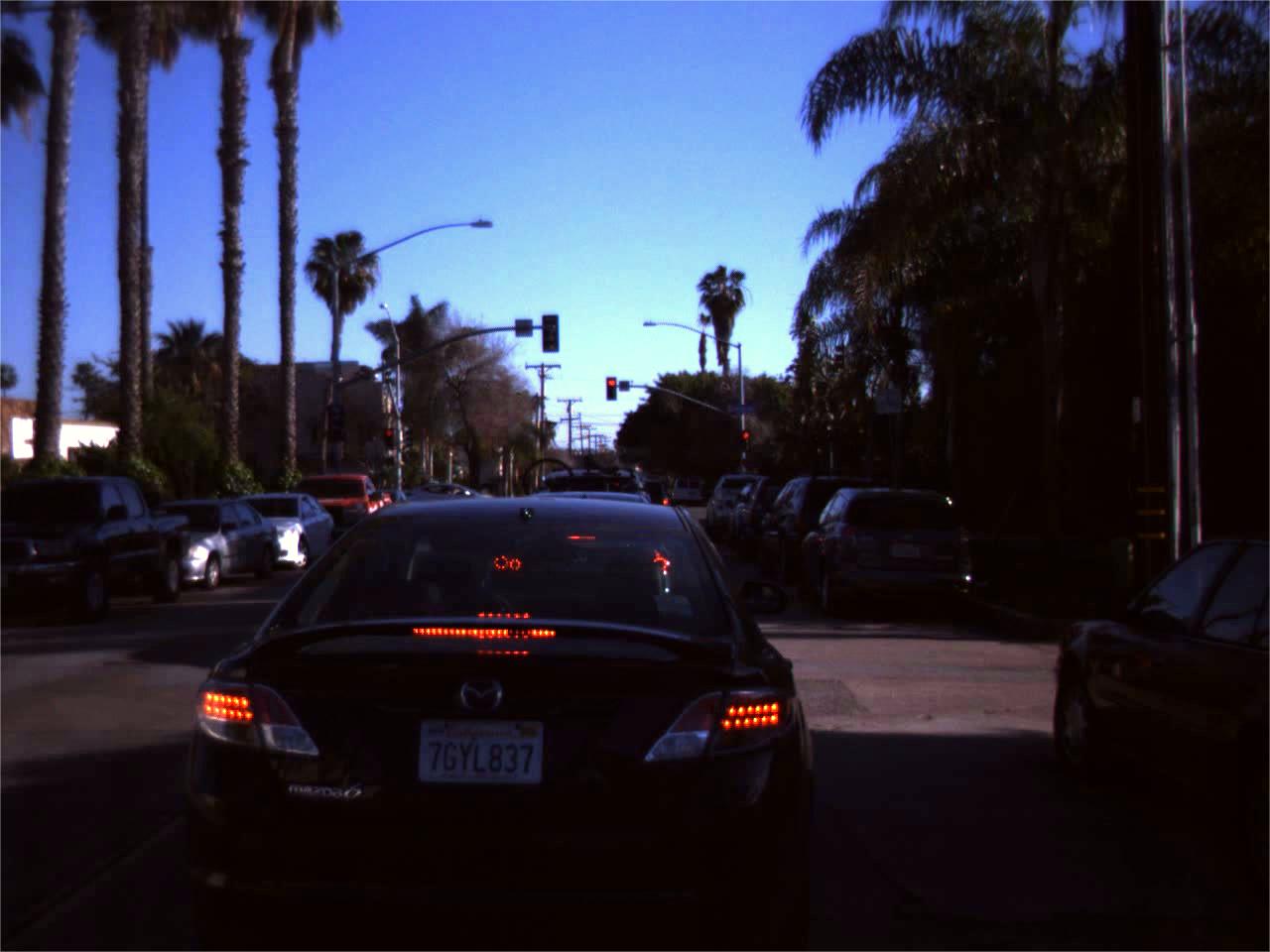}}
	\subfigure[RN]{
		\includegraphics[width=0.11\linewidth]{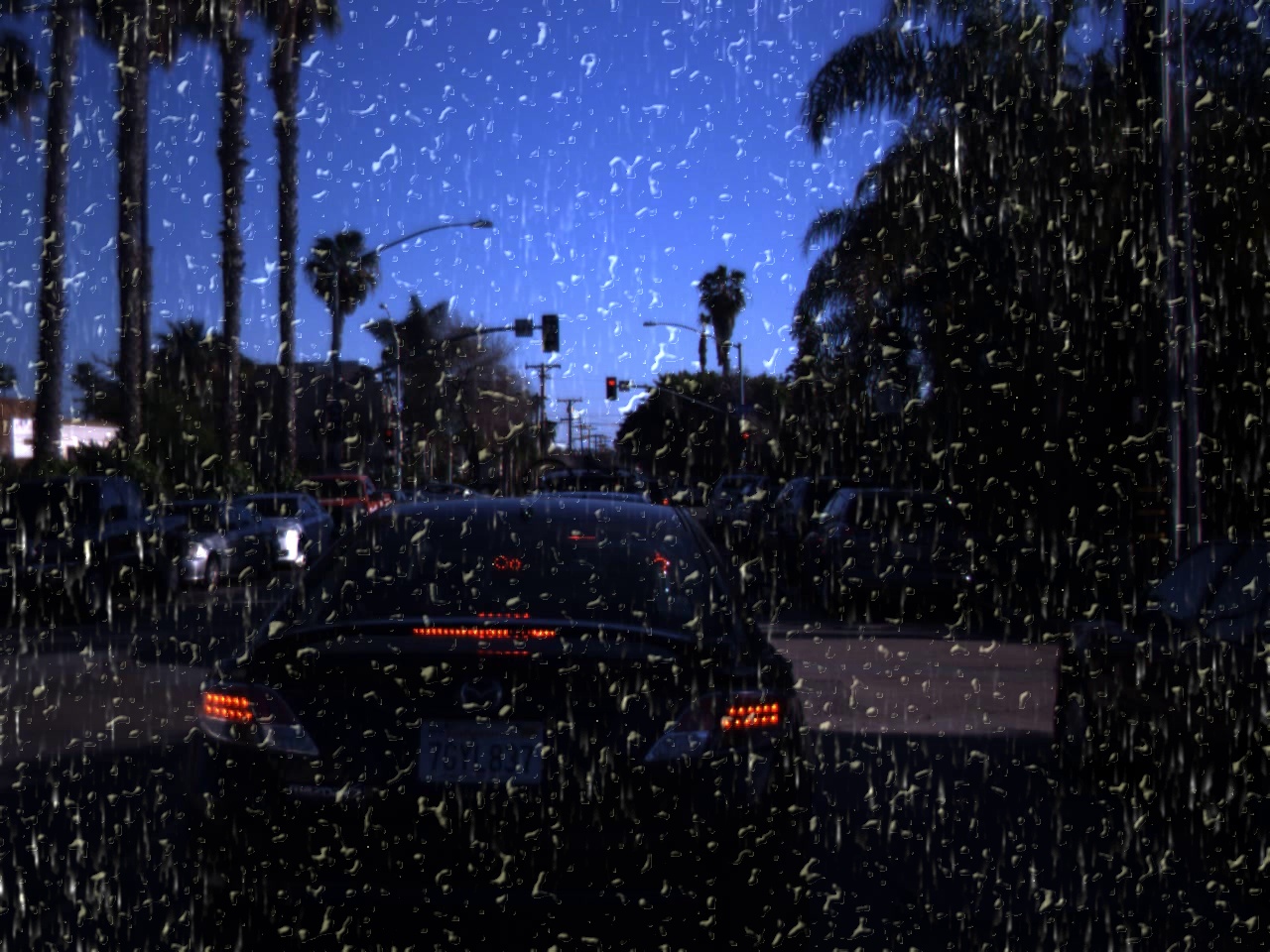}}
	\subfigure[OI]{
		\includegraphics[width=0.11\linewidth]{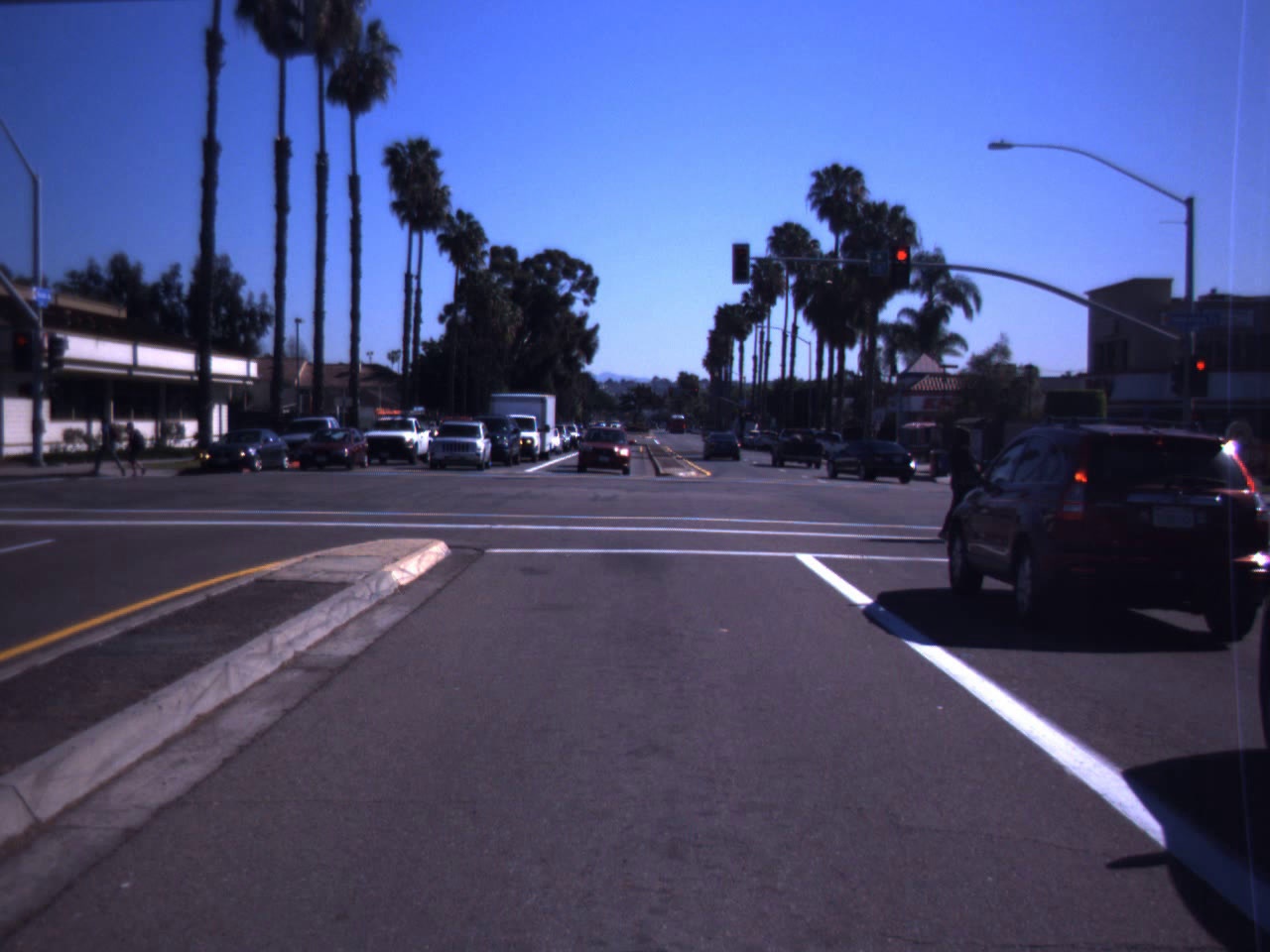}}
	\subfigure[SW]{
		\includegraphics[width=0.11\linewidth]{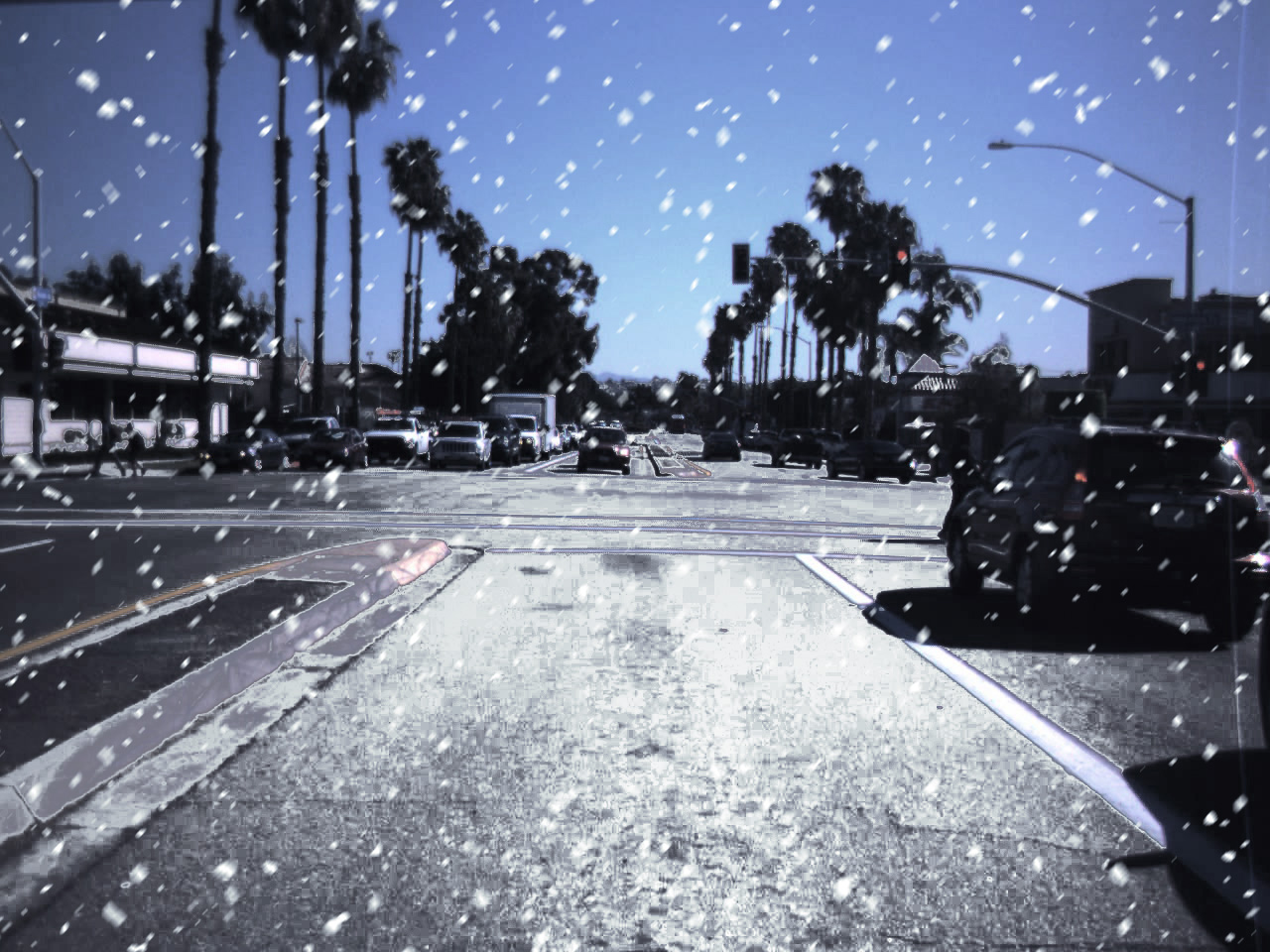}}
    \subfigure[OI]{
		\includegraphics[width=0.11\linewidth]{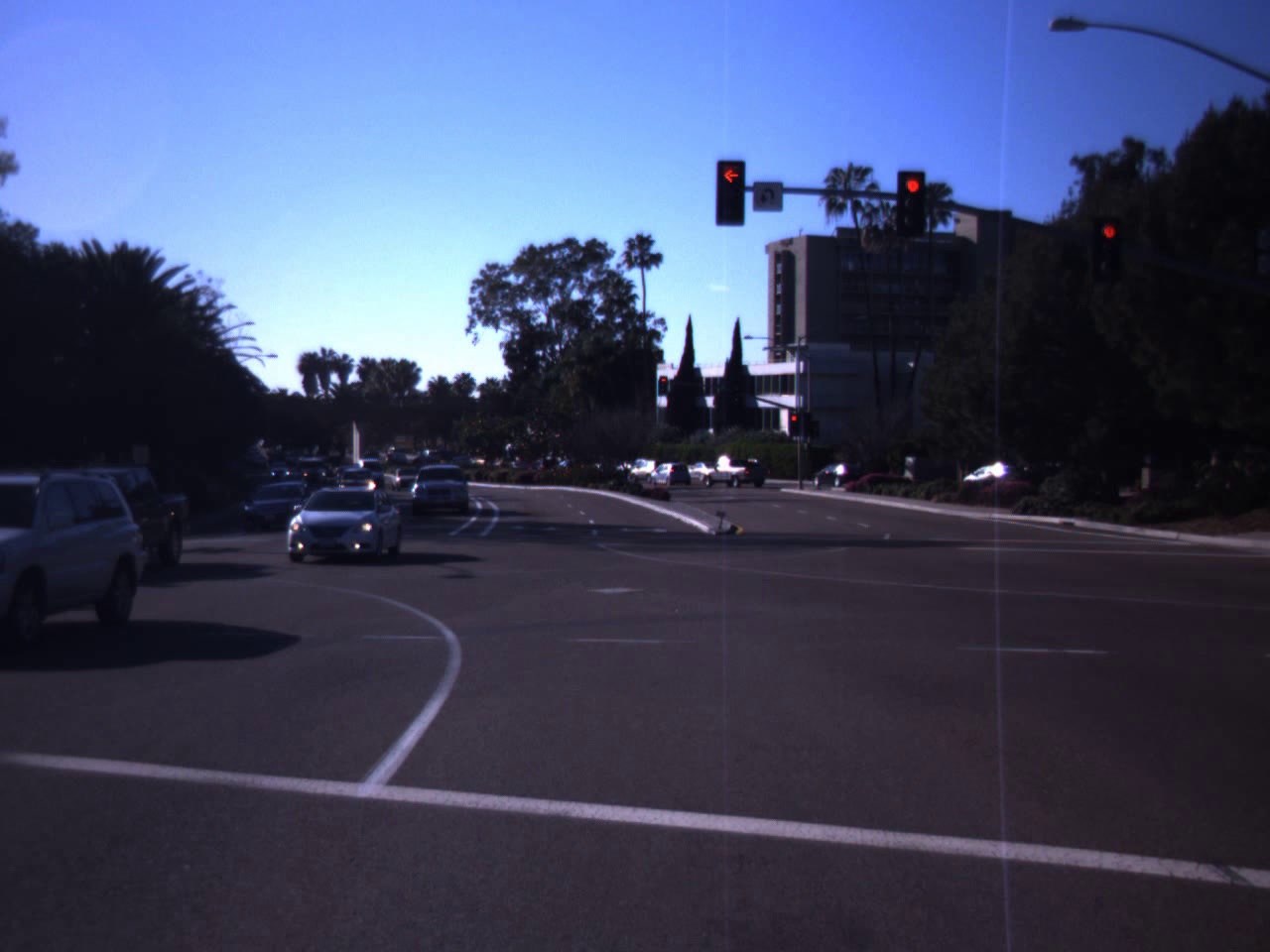}}
    \subfigure[FG]{
	    \includegraphics[width=0.11\linewidth]{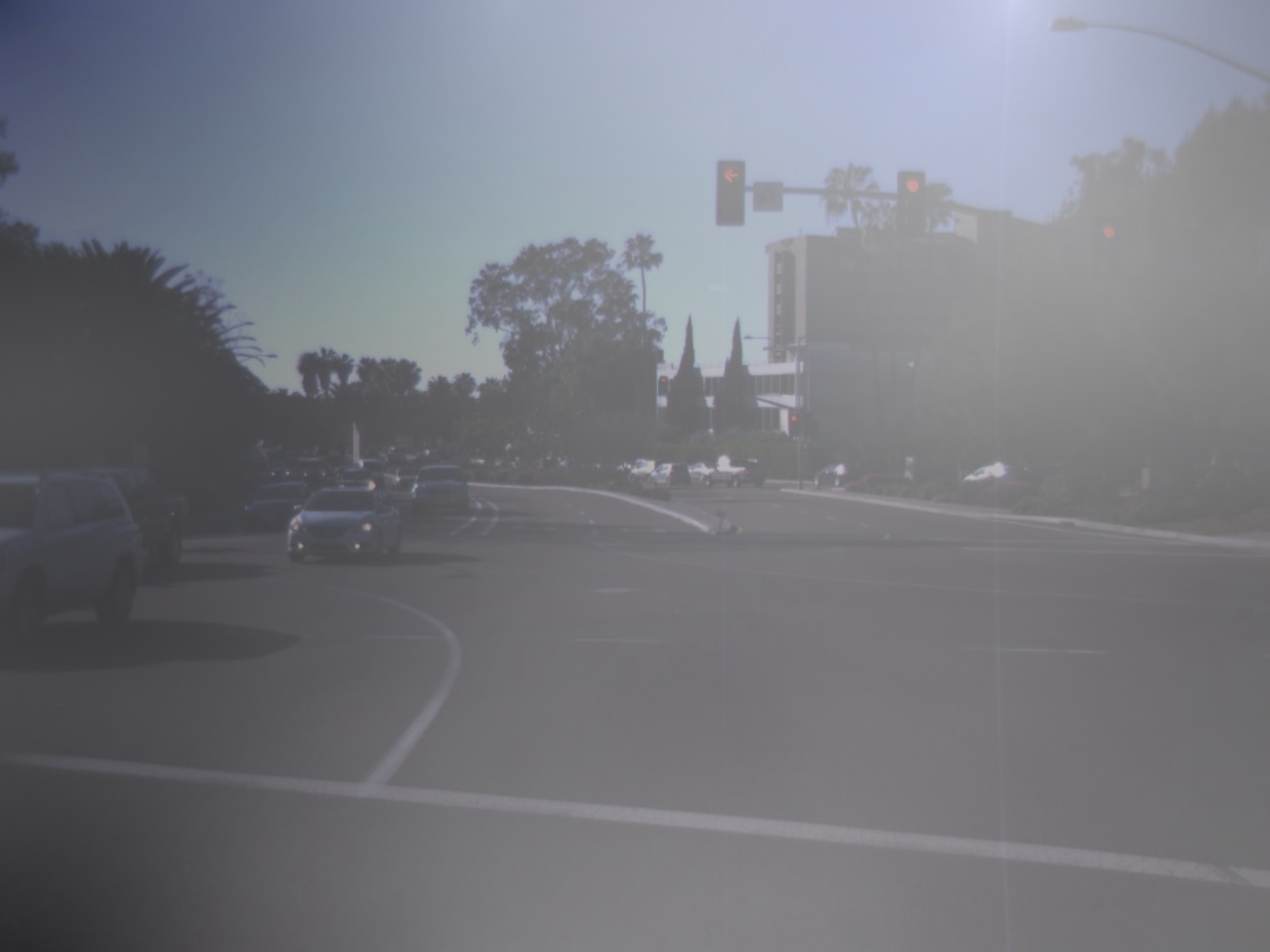}}
	\subfigure[OI]{
		\includegraphics[width=0.11\linewidth]{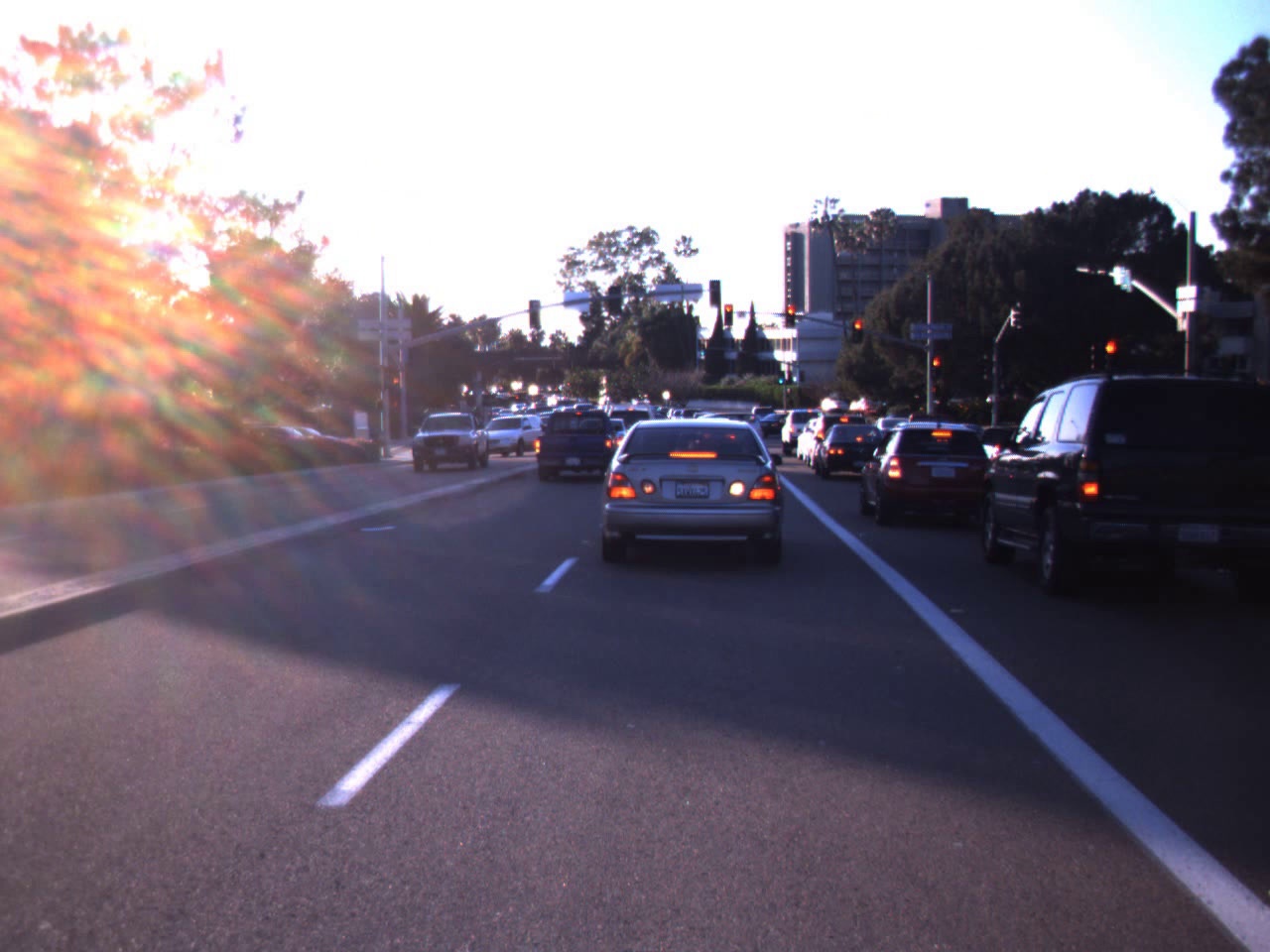}}
	\subfigure[LF]{
		\includegraphics[width=0.11\linewidth]{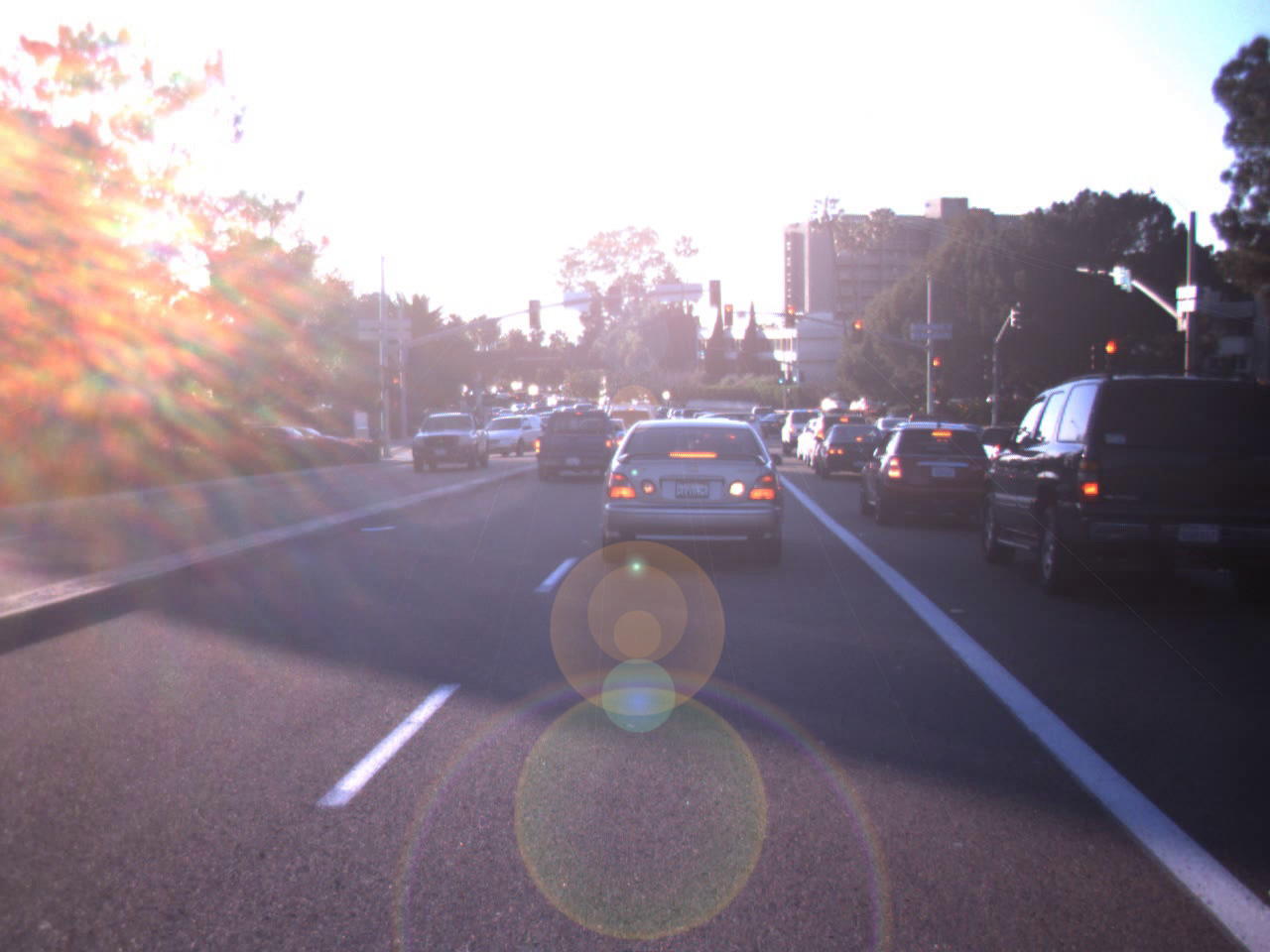}}
    
    \subfigure[OI]{
	    \includegraphics[width=0.11\linewidth]{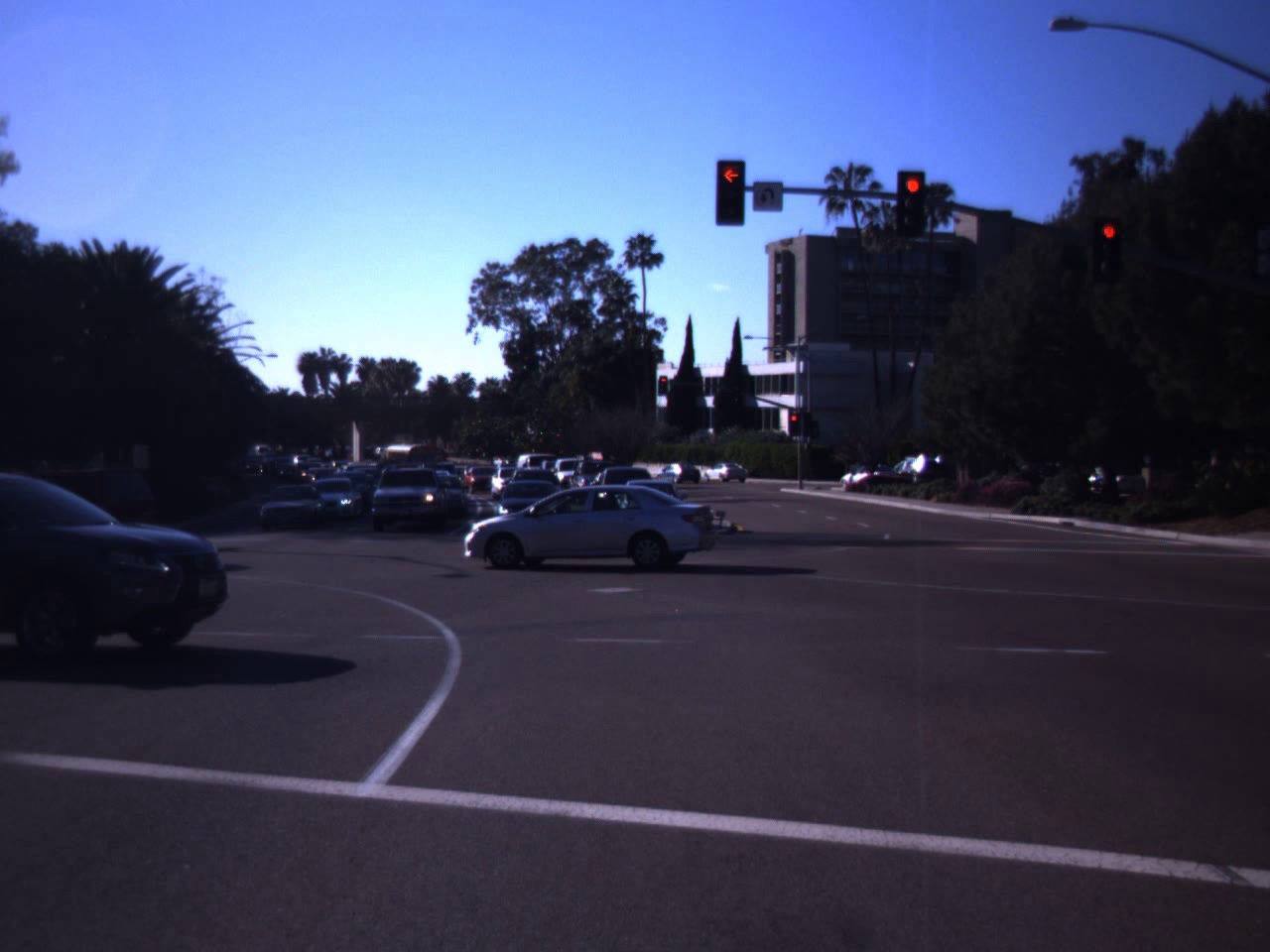}}
    \subfigure[OE]{
        \includegraphics[width=0.11\linewidth]{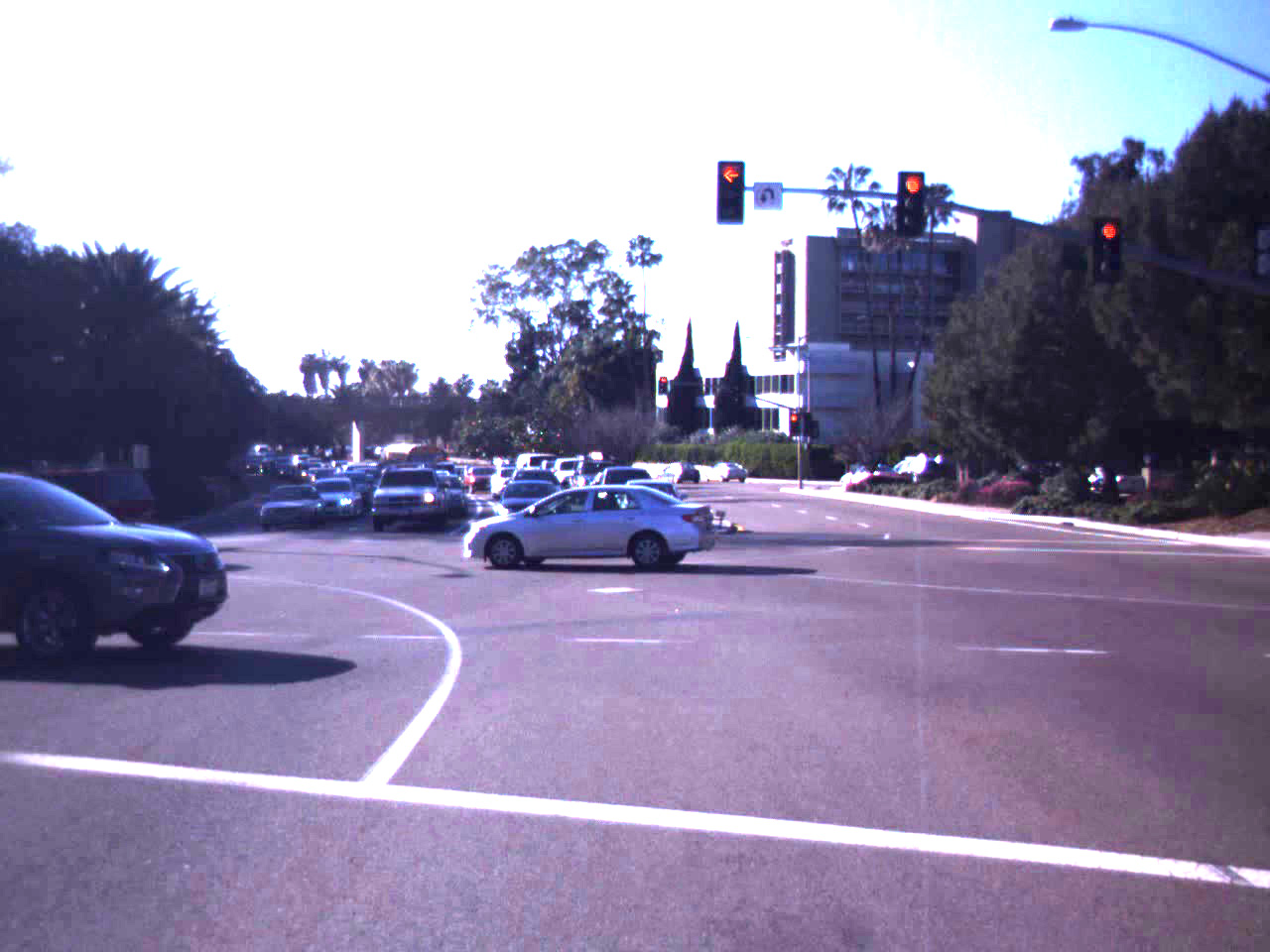}}
    \subfigure[OI]{
        \includegraphics[width=0.11\linewidth]{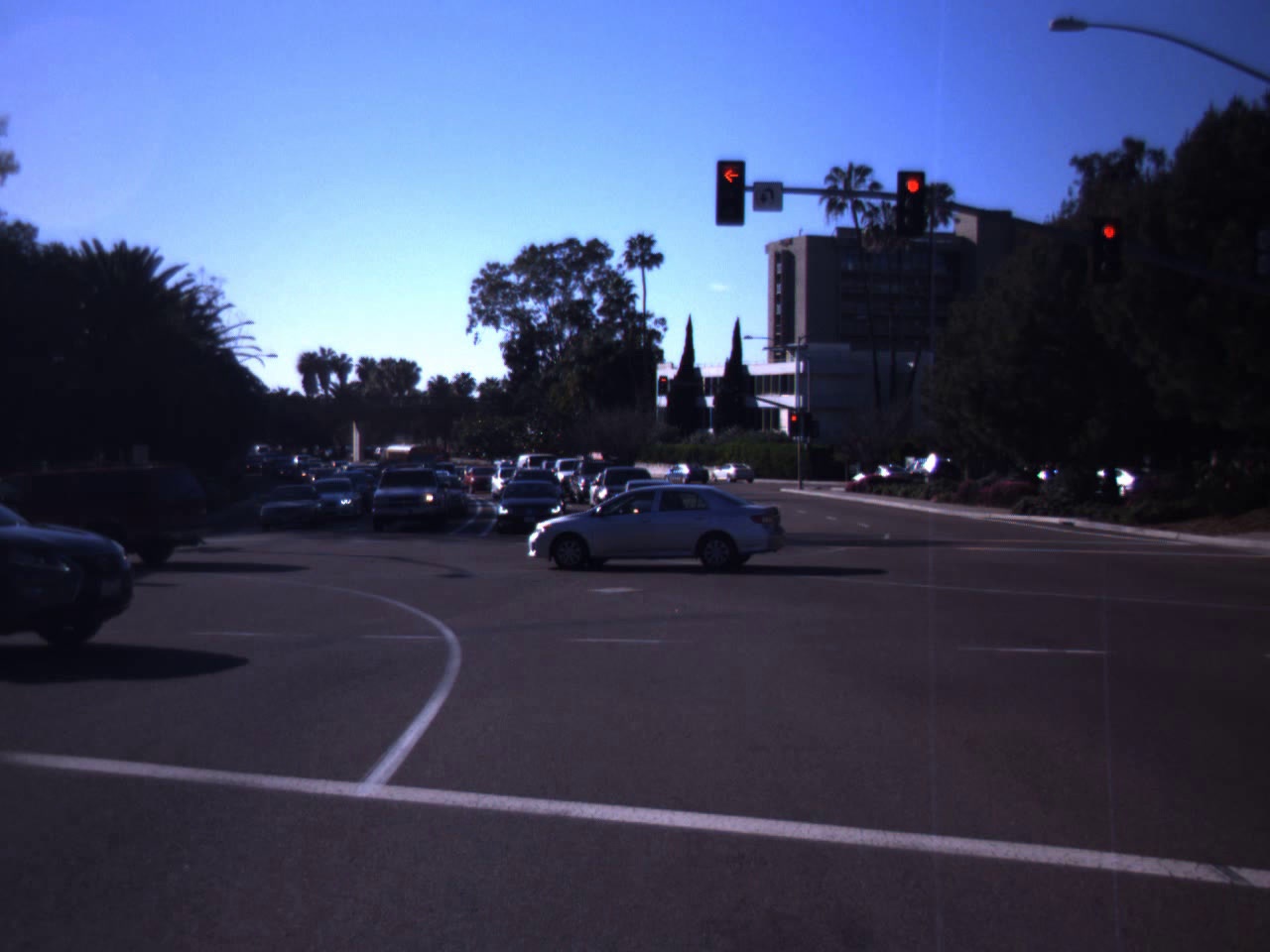}}
    \subfigure[UE]{
        \includegraphics[width=0.11\linewidth]{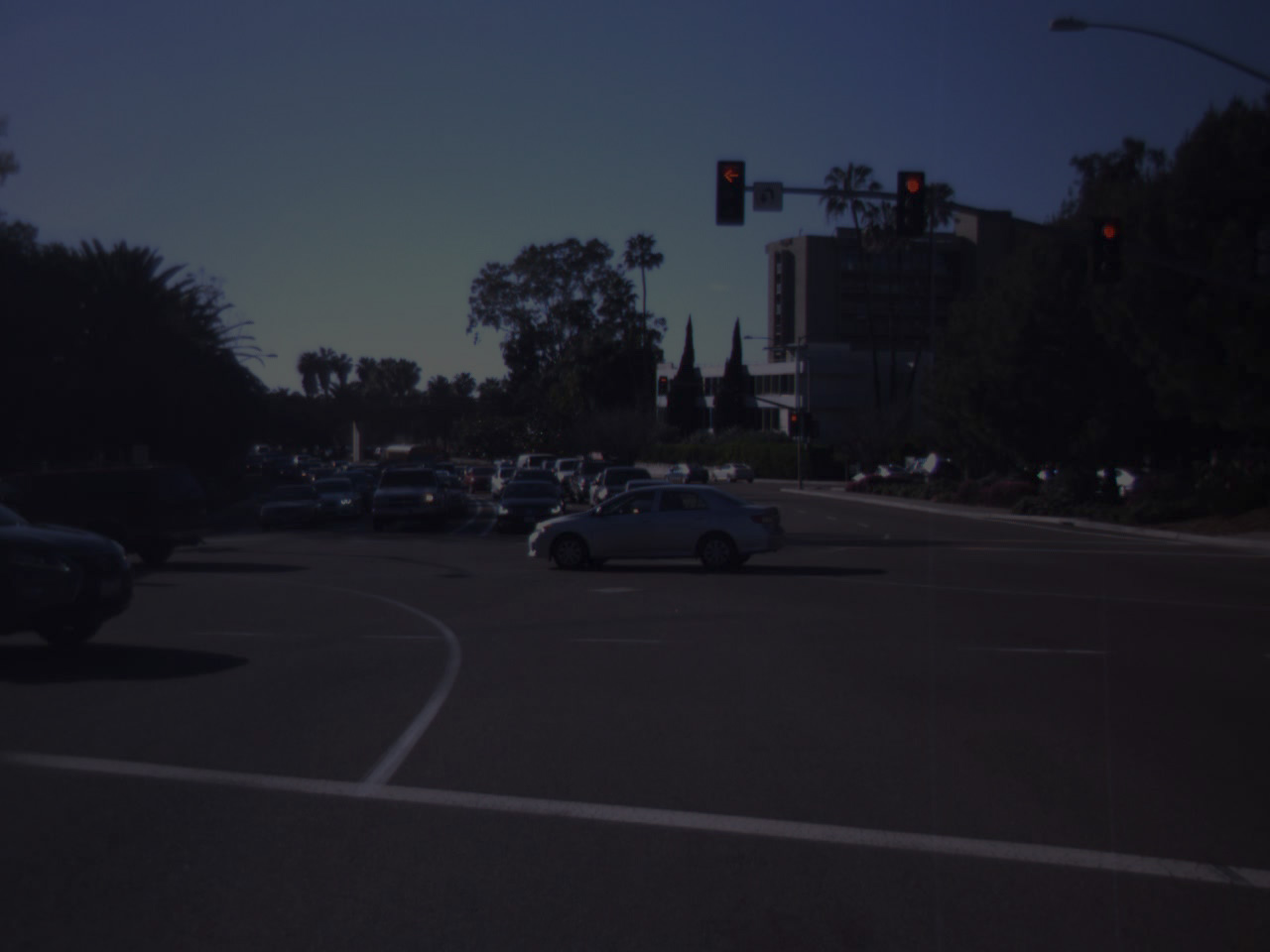}}	
    \subfigure[OI]{
        \includegraphics[width=0.11\linewidth]{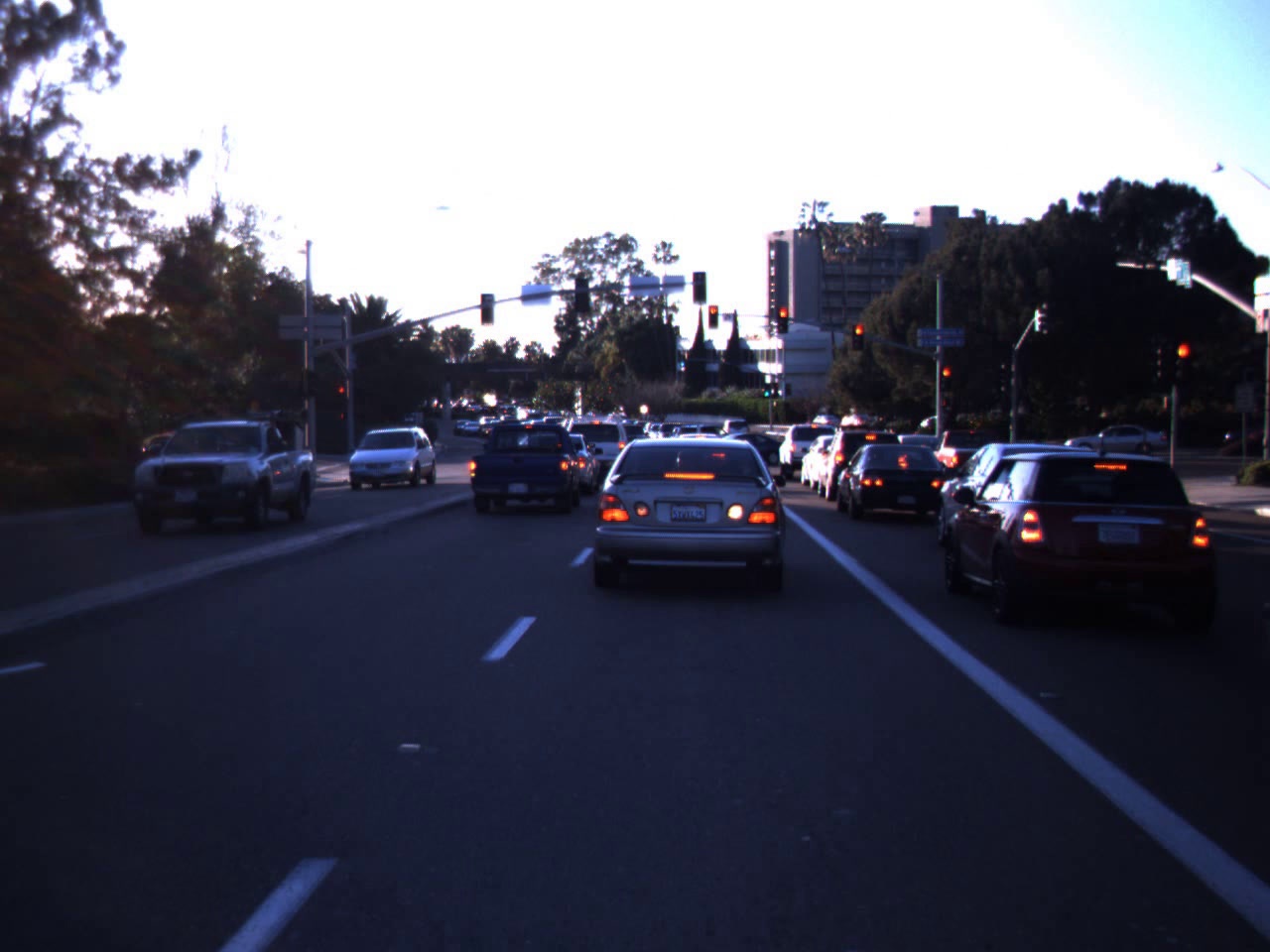}}
    \subfigure[MB]{
        \includegraphics[width=0.11\linewidth]{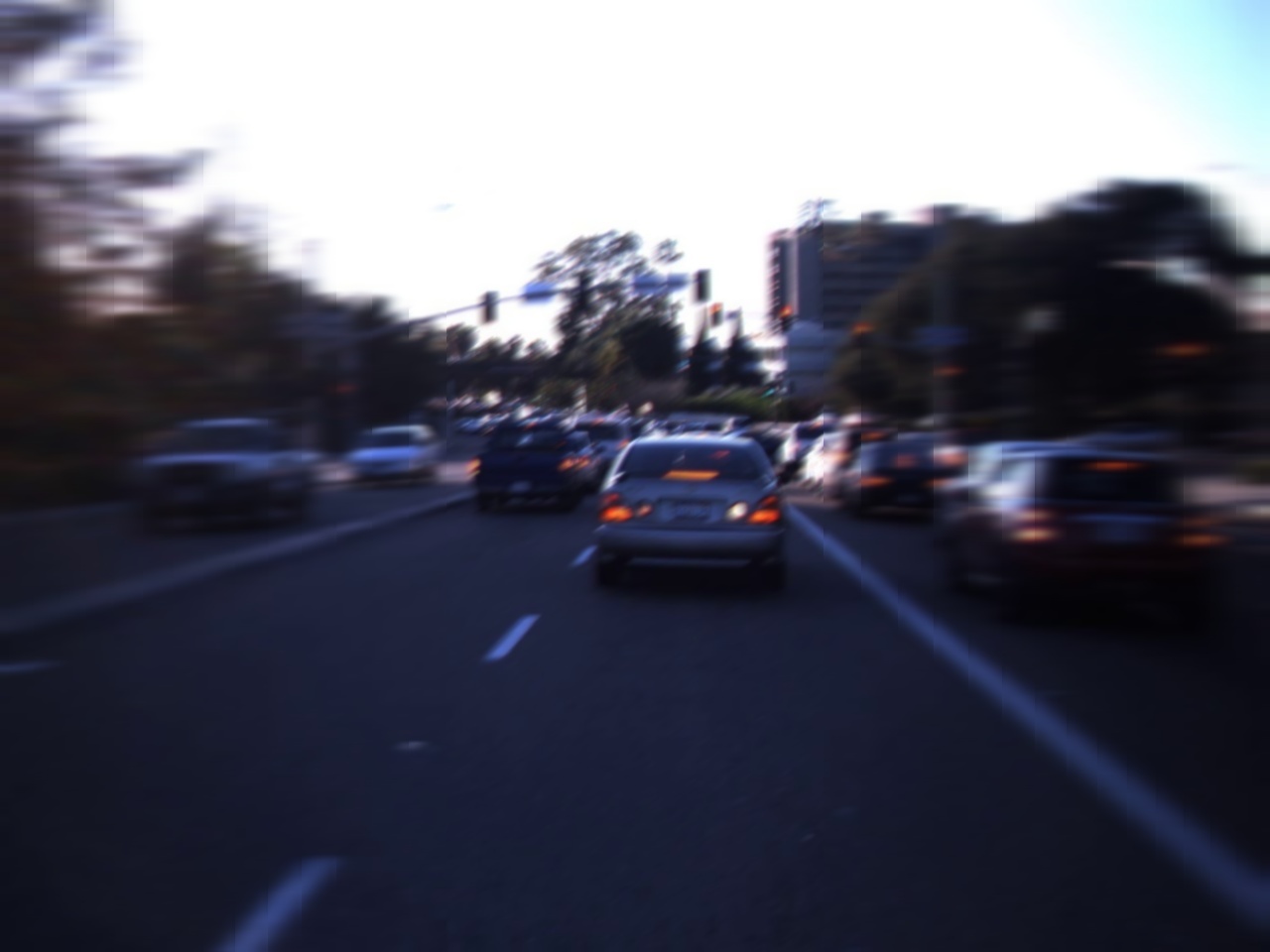}}
    \subfigure[OI]{
        \includegraphics[width=0.11\linewidth]{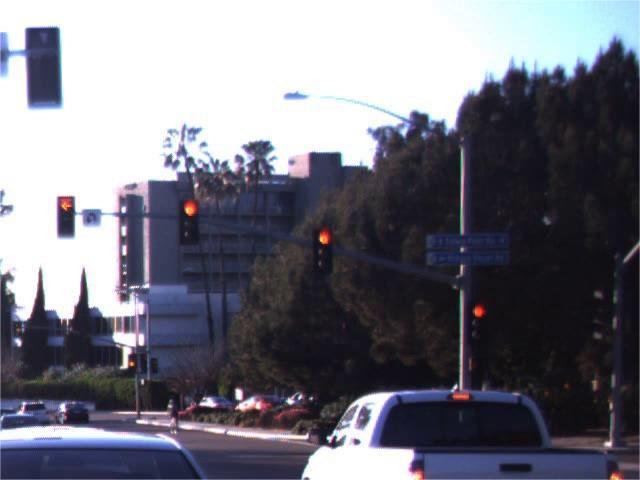}}
    \subfigure[CC]{
        \includegraphics[width=0.11\linewidth]{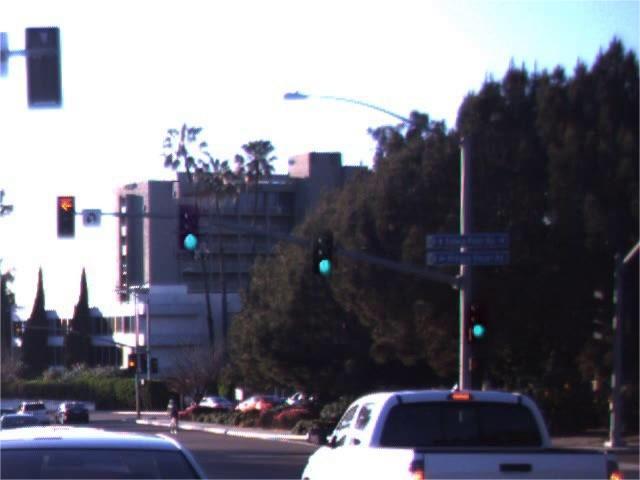}}
    
    \subfigure[OI]{
        \includegraphics[width=0.11\linewidth]{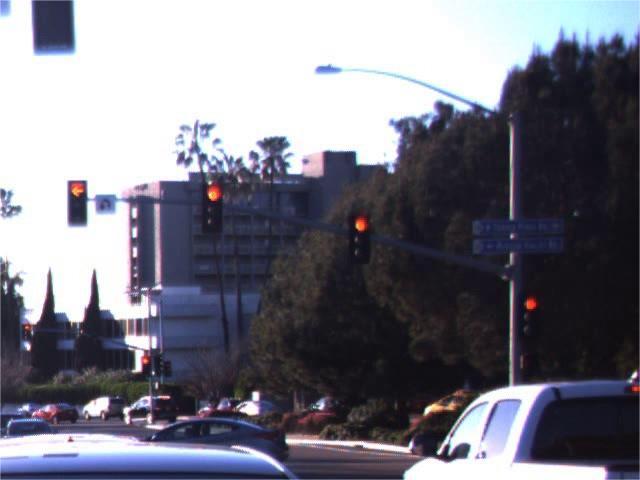}}     
    \subfigure[MP]{
        \includegraphics[width=0.11\linewidth]{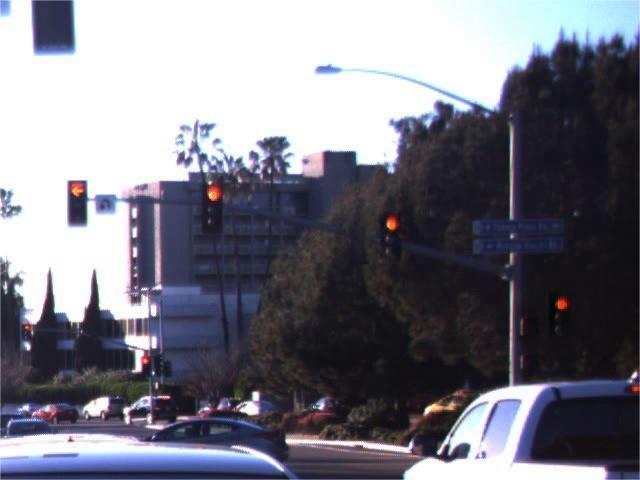}}
    \subfigure[OI]{
        \includegraphics[width=0.11\linewidth]{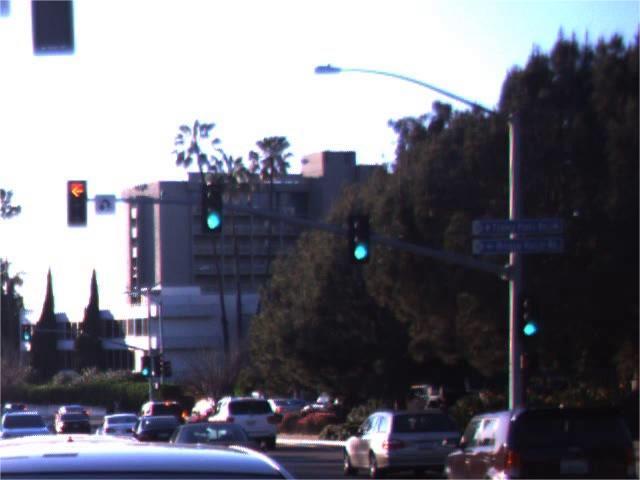}}
    \subfigure[AD]{
        \includegraphics[width=0.11\linewidth]{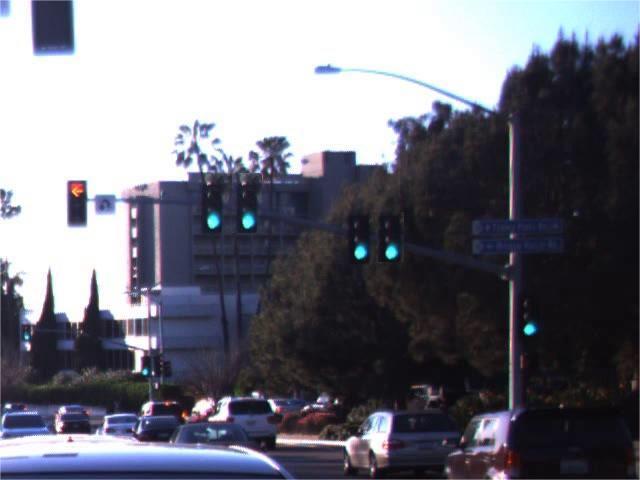}}   
    \subfigure[OI]{
	    \includegraphics[width=0.11\linewidth]{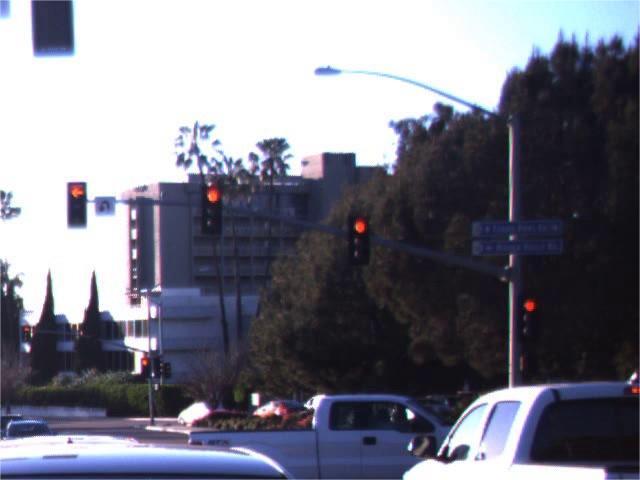}}
	\subfigure[RT]{
		\includegraphics[width=0.11\linewidth]{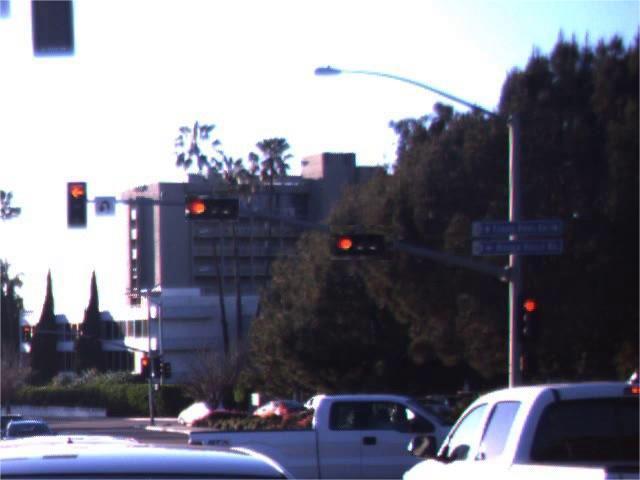}}
    \subfigure[OI]{
        \includegraphics[width=0.11\linewidth]{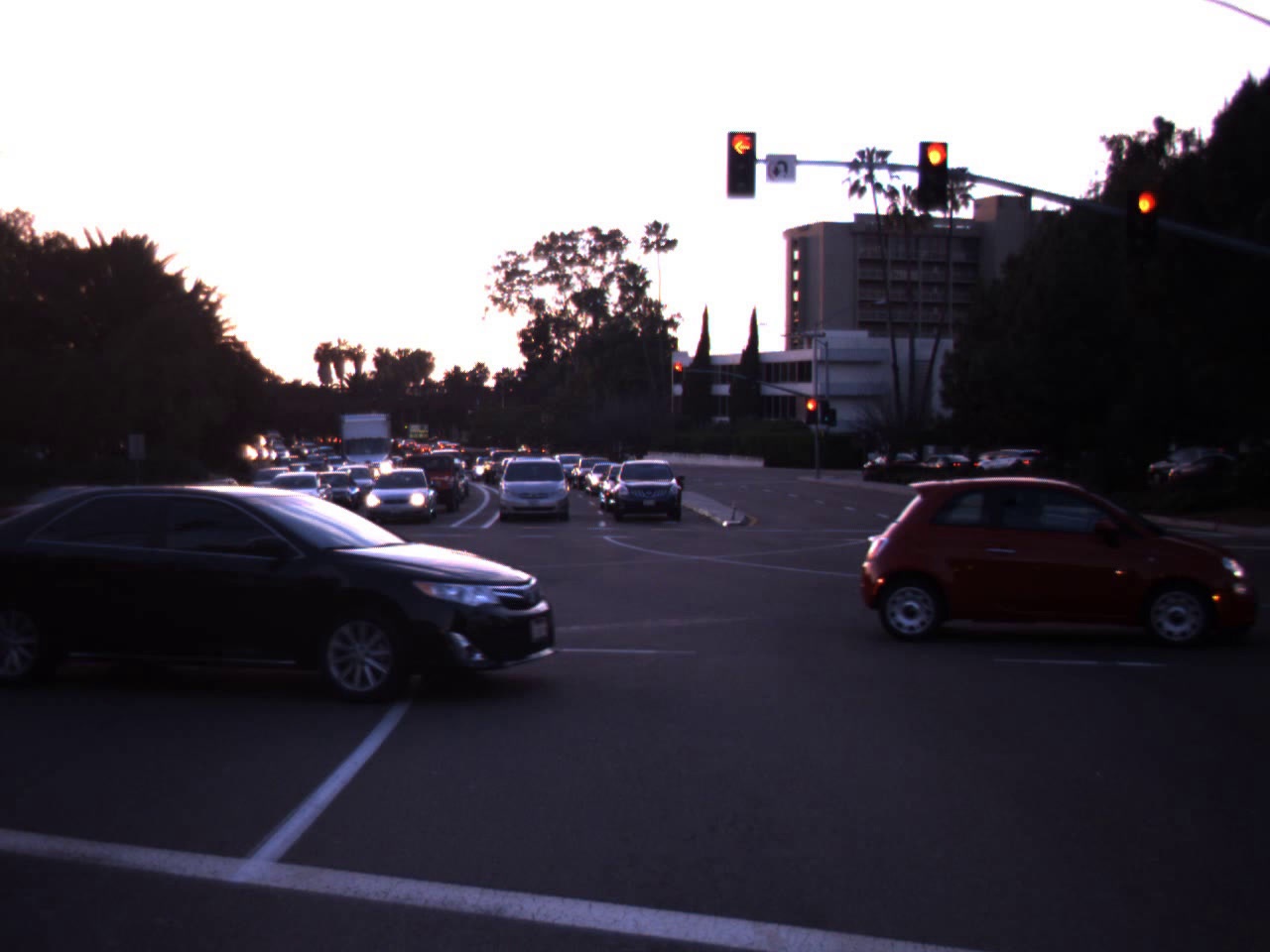}}
    \subfigure[SC]{
        \includegraphics[width=0.11\linewidth]{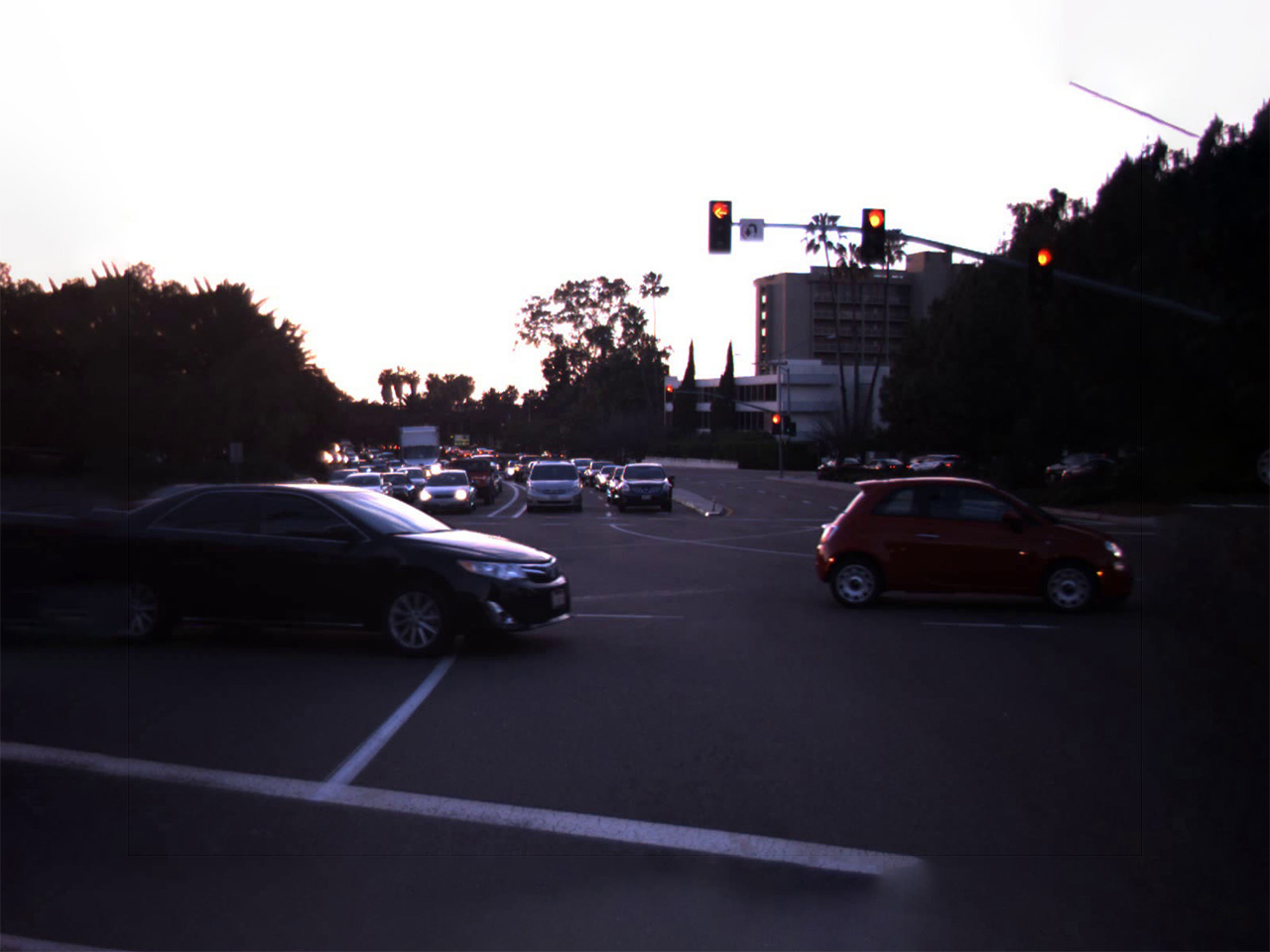}}   
%    \vspace{3pt}
    \caption{Sample Traffic Light Images Synthesized by Our Transformations}
    \label{fig:sample}
\end{figure*}

\textbf{Weather Transformations.} Autonomous vehicles drive~in different weather environments, which may partially block traffic~lights or reduce the visibility of traffic lights. For example, snowflakes~or~raindrops may happen to block traffic lights when the on-board camera captures the image. Therefore, weather environments may challenge the performance of traffic light detection. To simulate different weather environments, we design four transformations,~i.e.,~rain (RN), snow (SW), fog (FG) and lens flare (LF). We use the image~augmentation Python library \texttt{imgaug}~\cite{imgaug} to realize RN, SW~and~FG.~The basic idea is to generate and superimpose a layer of perturbation~on each pixel of the image to mimic weather conditions. Specifically,~for RN, we set the $drop\_size$ parameter to (0.1, 0.2) to control the size~of raindrops, and set the $speed$ parameter to (0.2, 0.3) to control the density of raindrops. For SW and FG, we set the $severity$ parameter to 2 to control the concentration of the fog or snow. Besides,~we~use Adobe Photoshop to realize LF, which mimics the dazzling light~effect captured by the camera. Specifically, we compose a new lens flare layer provided by Adobe Photoshop on the real-world traffic light image whose brightness and contrast are automatically~adjusted~\cite{lf-ps}. Fig.~\ref{fig:sample}(a-h) shows some sample traffic light images synthesized by our four weather transformations along with their original images (OI).

%\begin{center}
%    $aug = iaa.Rain(drop\_size=(0.10, 0.20), speed=(0.2, 0.3))$
%\end{center} 
%\begin{center}
%    $aug = iaa.imgcorruptlike.Fog(severity=2)$,
%\end{center}
%\begin{center}
%    $aug = iaa.imgcorruptlike.Snow(severity=2)$
%\end{center}

\textbf{Camera Transformations.} It is well known that the quality~of the images taken by on-board cameras may affect the performance of traffic light detection models. Due to various models of on-board cameras and their various ages and degrees of damage, the image captured by cameras may not have a good quality (e.g., overexposure or underexposure). Moreover, autonomous vehicles may~run~at a high speed~or~turn~at~intersections, which might blur the captured images to some extent. To simulate such camera effects, we develop three transformations, i.e., overexposure (OE), underexposure (UE) and motion blur (MB). Similar to weather transformations, we use the library \texttt{imgaug}~\cite{imgaug} to realize OE, UE and MB. For OE and UE, the image brightness is linearly adjusted by adjusting the lightness in the HSL color space, and we set the $severity$ parameter to 4 and~1 to control the degree of exposure for simulating overexposure and underexposure respectively. For MB, a filter that~uses~a~kernel~is~applied on an image through convolution, we set the kernel~size~parameter $k$ to 15 to control the degree of blur (i.e., a kernel size~of~15~$\times$ 15 pixels is used). Fig.~\ref{fig:sample}(i-n) illustrates some sample traffic light images synthesized by our three camera transformations along with their original images (OI).

%\begin{center}
%    $aug = iaa.imgcorruptlike.Brightness(severity=4)$,\\
%    $aug = iaa.MotionBlur(k=15)$
% \end{center}

\textbf{Traffic Light Transformations.} Traffic lights may have~different states and be placed at different positions in real~world. To enrich the diversity of positions and states of traffic~lights in images and ensure the synthesized traffic lights still follow existing regulations as much as possible, we design five traffic light transformations, i.e., changing~color~of~traffic lights (CC), moving position of traffic lights (MP), adding traffic lights (AD), rotating traffic lights (RT)~and scaling traffic lights~(SC).
 
\begin{itemize}[leftmargin=*]
\item CC is designed to change the color of traffic lights in an image. \ly{To ensure the semantic relationships among different traffic lights, we choose to change all the traffic lights to their opposite color (i.e., red lights turn to green, green lights turn to red, and yellow lights remain unchanged). Thus, the two traffic lights having the same state still represent the same direction after transformation. Specifically, we locate each of the red and green traffic light $t_{i}$ in $ T $ according to its bounding~box and extract it (i.e., the region corresponding to the bounding~box is blank after extraction). Then, we use content-aware patch~provided by Adobe Photoshop to fill the blank region.} Next,~we~use the HSV color space to change the hue of the extracted traffic~light, so that red becomes green and green becomes red, and~also~correspondingly change the traffic light state, i.e., $l_i.state$. Then, we flip the extracted traffic light upside down to ensure that the red light bulb is on the top or left of the traffic light. \ly{It is worth mentioning that some traffic lights also include arrows, and we also take these special cases into account during the transformation process to ensure that the stop-left traffic lights turn to go-left traffic lights. }Finally,~we~use image fusion algorithm (e.g., Poisson blending) to paste the transformed traffic light back to the original region. 

%The newly generated traffic light data set can be described as:
%    \begin{center}
%        $\hat{T}_{new}=\sum_{0\le i< ||T||  } f (\ C(\ P(t_{i}\ )\ )\ )+(\ \hat{T}-T\ )$
%    \end{center}
        
\item MP aims to move the position of traffic lights in an image. \ly{We first randomly select a subset of traffic lights~$\hat{T} \subseteq T $. For each $t_i \in \hat{T} \subseteq T$,  we extract $t_i$ and patch the~blank~region. Then, we paste $t_{i}$ to~a~new~position~by~a~horizontal offset $\delta$ (we set $\delta$ to the width of the bounding box so that the traffic lights will not move to unrealistic positions, e.g., overlapping with others).} Finally, we~modify the bounding box coordinates in $l_i$ as follows. %by Eq.~\ref{eq:bb1}.
\setlength{\abovedisplayskip}{4pt}
\setlength{\belowdisplayskip}{4pt}
\begin{equation}
% \begin{footnotesize}
\begin{aligned}\label{eq:bb1}
l_{i}.x_1 \doteq l_i.x_1+\delta,~l_{i}.x_2 \doteq l_i.x_2+\delta \nonumber
\end{aligned}
% \end{footnotesize}
\end{equation}

%     The newly generated traffic light data set can be described as:
%    \begin{center}
%        $\hat{T}_{new}=\sum_{0\le i< ||T||  } M(\ P(\ t_{i}\ )\ )+(\ \hat{T}-T\ )$
%    \end{center}
    
\item AD is designed to add traffic lights to an image.~The~number of newly added traffic lights is a random number between one and half of the original number of traffic lights.~The~basic idea~is~to~copy original traffic lights and paste them to the image. In that sense, AD is similarly implemented to MP.~The~difference~is~that here we do not patch the original traffic lights but keep them.
 
% The newly generated traffic light data set can be described as:   
%    \begin{center}
%        $\hat{T}_{new}=\sum_{0\le i< ||T||  } M(\ t_{i}\ )+ \hat{T} $
%    \end{center}
    
\item RT aims to rotate traffic lights in an image, so that horizontal~traffic lights become vertical and vertical traffic lights become horizontal. For each $t_i \in \hat{T} \subseteq T$,  we extract $t_i$ and~patch the blank region. Then, we determine the center of $t_i$, i.e., $(x_c, y_c)$ $= (\frac{l_i.x_1 + l_i.x_2}{2},\ \frac{l_i.y_1 + l_i.y_2}{2})$. In order to ensure that the red light bulb is on the top or left of the traffic light and the green light bulb is located at the bottom or right of the traffic light, we rotate $t_{i}$ by 90 degrees clockwise around the center if it is a horizontal traffic light, and we rotate it by 90 degrees counterclockwise if it is a vertical traffic light. \ly{Similar to CC, we also take traffic lights with arrows into consideration, e.g., we change the go-left traffic lights into go-straight traffic lights after RT transformation.} Finally, we paste the transformed traffic~light at the center of the original region, and correspondingly modify the bounding box coordinates in $l_i$ as follows. %by Eq.~\ref{eq:bb2}.
\setlength{\abovedisplayskip}{4pt}
\setlength{\belowdisplayskip}{4pt}
\begin{equation}
% \begin{footnotesize}
\begin{aligned}\label{eq:bb2}
l_{i}.x_1 \doteq x_{c}-\frac{l_i.y_2-l_i.y_1}{2},~l_{i}.y_1 \doteq y_{c}-\frac{l_i.x_2-l_i.x_1}{2}\\\nonumber
l_{i}.x_2 \doteq x_{c}+\frac{l_i.y_2-l_i.y_1}{2},~l_{i}.y_2 \doteq y_{c}+\frac{l_i.x_2-l_i.x_1}{2}
\end{aligned}
% \end{footnotesize}
\end{equation}
    
%    And the newly generated traffic light data set can be described as:
%    \begin{center}
%        $\hat{T}_{new}=\sum_{0\le i< ||T||  } R(\ P(\ t_{i}\ )\ )+(\ \hat{T}-T\ )$
%    \end{center}

\item SC is designed to scale the size of traffic lights to mimic~the effect of taking the image from a greater distance.~We~implement SC with the help of the batch processing function of Adobe Photoshop. First, we import each image into Adobe Photoshop~and~expand the canvas according to its original size (e.g., increase the width by 320 pixels and the height by 180 pixels for~a~16:9~image). Then, we fill the expanded region with uni-color and patch it using the image inpainting technique in Adobe Photoshop.~Finally, we scale the new image to its original size. Due~to~the~scaling, the bounding box of each traffic light is scaled correspondingly as follows, where $w$ and $h$ denote the original width~and~height of the bounding box, and $(x_c, y_c)$ denotes the center.
\setlength{\abovedisplayskip}{4pt}
\setlength{\belowdisplayskip}{4pt}
\begin{equation}
% \begin{tiny}
\begin{aligned}\label{eq:bb3}
    l_{i}.x_1 \doteq x_{c}-(\frac{l_i.x_2-l_i.x_1}{2}\times\frac{w}{w+320}),~l_{i}.y_1 \doteq y_{c}-(\frac{l_i.y_2-l_i.y_1}{2}\times\frac{h}{h+180})\\\nonumber
    l_{i}.x_2 \doteq x_{c}+(\frac{l_i.x_2-l_i.x_1}{2}\times\frac{w}{w+320}),~l_{i}.y_2 \doteq y_{c}+(\frac{l_i.y_2-l_i.y_1}{2}\times\frac{h}{h+180})
\end{aligned}
% \end{tiny}
\end{equation}

\end{itemize}

Fig.~\ref{fig:sample}(o-x) reports some sample traffic light images synthesized by our five traffic light transformations along with their original images (OI).

\section{Evaluation}

We first introduce the research questions, then elaborate our evaluation setup, and finally present our results. 

% original training datasets
% original testing datasets
% augmented training datasets
% augmented testing datasets.
% improved augmented training datasets
% improved augmented testing datasets.

% original models
% retrained models using augmented training datasets
% retrained models using improved augmented training datasets

\subsection{Research Questions}

% \tool is designed to improve performance of traffic light detection in autonomous driving systems using data augmentation. Specifically, we transform traffic light images from two datasets based on three families of metamorphic relations (see Sec. \ref{sec:approach}). Then, we use augmented traffic light images to retrain four popular traffic light detection models. 

We design the following four research questions to evaluate~the~effectiveness and efficiency of \tool.

\begin{itemize}[leftmargin=*]
    \item \textbf{RQ1 Robustness Evaluation}: How effective are our augmented traffic light images in identifying erroneous behaviors of existing traffic light detection models?% using our metamorphic relations?
    \item \textbf{RQ2 Retraining Evaluation}: How effective are our~augmented traffic light images in improving the performance~of existing traffic light detection models?% via retraining?
    \item \textbf{RQ3 Efficiency Evaluation}: How is the time overhead of \tool in synthesizing images and retraining models?
    \item \textbf{RQ4 Naturalness Evaluation}: How is the naturalness~of~augmented traffic light images, and does it affect the performance of existing traffic light detection models?
\end{itemize}

\textbf{RQ1} is designed to detect potential erroneous behaviors~in existing traffic light detection models by investigating the~performance difference between augmented testing data and real-world testing data against the models trained from real-world training~data.~\textbf{RQ2} aims to analyze the potential performance improvement~after retraining the models with~augmented~training~data~resulted~from~twelve transformations. Specifically,~we compare the performance difference for both real-world testing data and augmented testing~data. \textbf{RQ3} is designed to analyze the time cost of applying transformations as well as the time cost of retraining models. We aim~to~investigate the practical cost of \tool's effectiveness in detecting erroneous behaviors and improving model performance. \textbf{RQ4} aims to assess the naturalness of augmented images by manually inspecting and excluding ``unnatural'' images in order to make the rest of them resemble real-world images. We further investigate how the cleaned augmented data affect the model performance.

% We retrain the models using the improved augmented training datasets. Then, wea observe the accuracy of the original models, models retrained using augmented training datasets and models retrained using improved augmented training datasets.

\subsection{Evaluation Setup}

%We elaborate evaluation setup including \textbf{Datasets}, \textbf{Models} and \textbf{Environment}.

% We run our evaluation with two traffic light datasets and~four state-of-the-art traffic light detection models.

% O, rn+, rn- 
% \footnote{https://www.kaggle.com/datasets/mbornoe/LISA-traffic-light-dataset}
% \footnote{https://hci.iwr.uni-heidelberg.de/content/bosch-small-traffic-lights-dataset}

\textbf{Datasets.} We select two datasets, \textit{LISA}~\cite{jensen2016vision, philipsen2015traffic}~and \textit{Bosch} \cite{behrendt2017deep}. They are two of the most popular datasets~in~autonomous driving research area. \textit{LISA} consists of continuous testing~and training video sequences collected in California,~USA. %~The~sequences are captured by a stereo camera mounted on the roof of a vehicle driving under both night and daytime with varying light and weather conditions. 
The~original \textit{LISA} dataset totals \todo{43,007} images and \todo{113,888} annotated traffic lights. The original \textit{Bosch} dataset contains \todo{13,427} images and about \todo{24,000} annotated traffic lights. The annotations~include bounding boxes of traffic lights and~the~current state~of each traffic light. We filter monochrome images~from~the~original \textit{Bosch} dataset. We also remove duplicated images from both \textit{LISA} and \textit{Bosch} dataset. \ly{Moreover, we skip images with zero traffic light according to annotated data during the traffic light transformations process, but keep these images that do not contain traffic lights for training and testing.} Finally, we obtain preprocessed original datasets with \todo{36,265} images and \todo{109,475} annotated traffic lights in \textit{LISA} as well as \todo{10,300}~images~and \todo{24,242} annotated traffic lights~in~\textit{Bosch}.~Each~dataset~is~split into training data, validation data and testing data by 4:1:1. We use \texttt{O} to represent real-world datasets before augmentation.

To obtain augmented datasets for \textbf{RQ1}, \textbf{RQ2} and \textbf{RQ3}, we apply our twelve transformations on \textit{LISA} and \textit{Bosch}.  \ly{We generate the training and testing datasets for the 12 transformations from the same training and testing images of \textit{LISA} and \textit{Bosch} datasets respectively. As a result,~we~obtain 24 augmented datasets.} We represent each augmented dataset~using its corresponding transformation name and~a~`+'~symbol; e.g., \todo{\texttt{RN+} represents~the~augmented dataset after applying our RN transformation}. Besides,~to~support~naturalness evaluation in \textbf{RQ4}, we recruit three graduate students to manually inspect the ``naturalness'' of each transformed~image in 12 augmented datasets resulted~from~\textit{LISA}. To~reduce~the manual effort, we do not analyze the smaller dataset \textit{Bosch}. Two~participants first conduct a pilot labeling~on~randomly~sampled \todo{6,000} images to determine whether the transformed image is ``natural''. When there is a conflict between~the~two~participants, the third participant is involved to have a group~discussion and reach agreements. The process takes \todo{3} more rounds until the Cohen Kappa coefficient reaches \todo{0.845}. Finally, the two participants go through~the~rest~augmented data to filter ``unnatural'' images. The whole~process takes \todo{2} human-months. We represent each cleaned~augmented~dataset after manual cleaning using its corresponding transformation name and a `-' symbol; e.g., \todo{\texttt{RN-} represents the cleaned augmented dataset after applying our RN transformation and manual cleaning}.

% model, model+, model-
% \footnote{https://github.com/ultralytics/yolov5}
% \footnote{https://github.com/Megvii-BaseDetection/YOLOX}
% \footnote{https://github.com/ShaoqingRen/faster\_rcnn}
% \footnote{https://github.com/amdegroot/ssd.pytorch}
\textbf{Models.} We choose four state-of-the-art traffic light detection models, i.e., \textit{YOLOv5}, \textit{YOLOX}, \textit{Faster R-CNN} and~\textit{SSD}.~These models are widely used in autonomous driving research~\cite{pan2019traffic, kim2018deep, muller2018detecting, behrendt2017deep, li2021improved}. Briefly, \textit{YOLO} model and its generations~(e.g., \textit{YOLOv3}, \textit{YOLOv5} and \textit{YOLOv8})~\cite{redmon2017yolo9000} are cutting-edge~and~state-of-the-art models~in~a~variety of tasks, including object~detection, image segmentation and image classification. We choose \textit{YOLOv5} because it is mature~and~stable. Further, \textit{YOLOX}~is~a popular variant of \textit{YOLO}, redesigned for anchor-free as well as other improvements for better performance. \textit{Faster R-CNN}~\cite{ren2015faster} is an object detection framework based on deep convolutional networks including a region proposal network and an object detection network. Both networks are trained for sharing convolutional layers for fast testing. \textit{SSD}~\cite{liu2016ssd} is implemented on a~single~shot~multi-box~detector.

For \textbf{RQ1}, we obtain eight original models by training the four models \textit{YOLOv5}, \textit{YOLOX}, \textit{Faster R-CNN} and \textit{SSD} with the two original training datasets from \textit{LISA} and \textit{Bosch}. For \textbf{RQ2} and \textbf{RQ3}, we obtain 96 retrained models by retraining \textit{YOLOv5}, \textit{YOLOX}, \textit{Faster~R-CNN} and \textit{SSD} with the 24 augmented training datasets. Specifically, to reduce the retraining time cost, we randomly choose~20\%~of~each augmented training dataset to merge into the original training~dataset respectively. Notice that existing test case selection approaches \cite{Ma2021test} can be used here to further improve the retraining effectiveness. Moreover, to investigate the effect of putting all transformations together, we obtain another eight retrained models by including all the above 20\% of the 12 augmented training datasets into~the~original training dataset. For \textbf{RQ4}, we obtain 24 retrained models by retraining \textit{YOLOv5} and \textit{YOLOX} with the 12 cleaned augmented training datasets from~\textit{LISA}.~To~align with the retraining in \textbf{RQ2} and \textbf{RQ3}, we first use the same 20\% of augmented training dataset and then add extra cleaned augmented training samples if some of them are cleaned for ``unnaturalness''. \todo{Here we do not retrain \textit{Faster~R-CNN} and \textit{SSD} because their training time is much longer than \textit{YOLOv5} and \textit{YOLOX}, and their detection performance is much lower than \textit{YOLOv5} and \textit{YOLOX}.}

We distinguish between original and retrained~models using the following notions. First, \texttt{M-O} represents an original~model where \texttt{M} is a notion for model and \texttt{O} denotes the original~training dataset. Second, \texttt{M-$\tau$+} denotes a retrained model~using~augmented training dataset by a transformation~$\tau \in \mathbb{W} \cup \mathbb{C}  \cup \mathbb{L}$. For~example, \texttt{M-RN+} denotes the retrained model using~augmented training~data-set by our \textit{RN} transformation. Third,~\texttt{M-$\tau$-} denotes a retrained model using cleaned augmented training dataset by a transformation $\tau \in \mathbb{W} \cup \mathbb{C} \cup \mathbb{L}$ after manual cleaning. For example,~\texttt{M-RN-}  denotes the retrained model using cleaned augmented training~dataset~by our \textit{RN} transformation. Fourth, we also use concrete model's names to represent \texttt{M}.~For~example, \texttt{YOLOX-O} denotes the original \textit{YOLOX} model using the original training dataset;~and~\texttt{YOLOX+} (resp. \texttt{YOLOX-}) denotes the retrained \textit{YOLOX} model using (resp. cleaned) augmented training dataset without distinguish transformations.

\textbf{Metric and Environment.} We use mAP (a value between~0 and 1), as defined in Sec.~\ref{sec:mr}, to measure model performance. We conduct the experiments on an Ubuntu 20.04.4 LTS server with 4 NVIDIA GeForce RTX 3090 GPUs, Intel Core i9-10980XE CPU with 3.00GHz processor and 128GB memory.

% !TeX root = ../main.tex

\subsection{Robustness Evaluation (RQ1)} 

\begin{figure*}[!t]
    \centering
  % \subfigbottomskip=2pt 
  %   \subfigcapskip=-4pt
    \subfigure[Results on the \textit{LISA} Dataset]{
		        \includegraphics[width=0.47\linewidth]{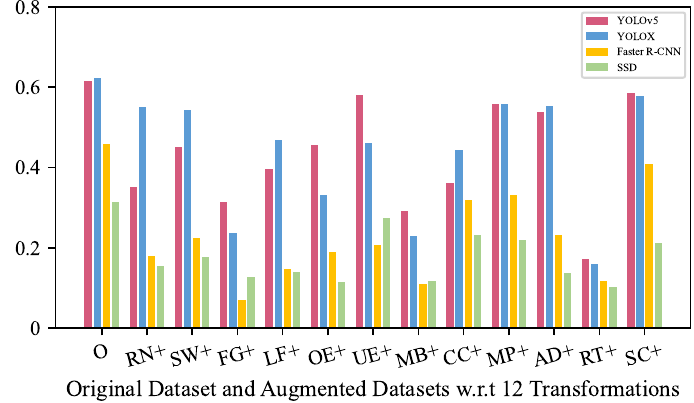}\label{fig:rq1_1}}
    \subfigure[Results on the \textit{Bosch} Dataset]{
            \includegraphics[width=0.47\linewidth]{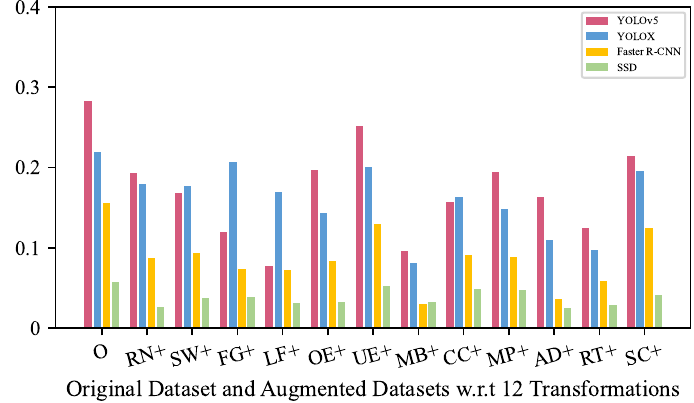}\label{fig:rq1_2}}
    % \vspace{-5pt}
    \caption{mAP Comparison of the Original Models between Original and Augmented Testing Datasets}\label{fig:rq1}
\end{figure*}

\begin{figure*}[!t]
    \centering  
    % \vspace{-5pt}
%  \subfigbottomskip=0pt 
%  \subfigcapskip=0pt
    \subfigure[OI]{
      \includegraphics[width=0.15\linewidth]{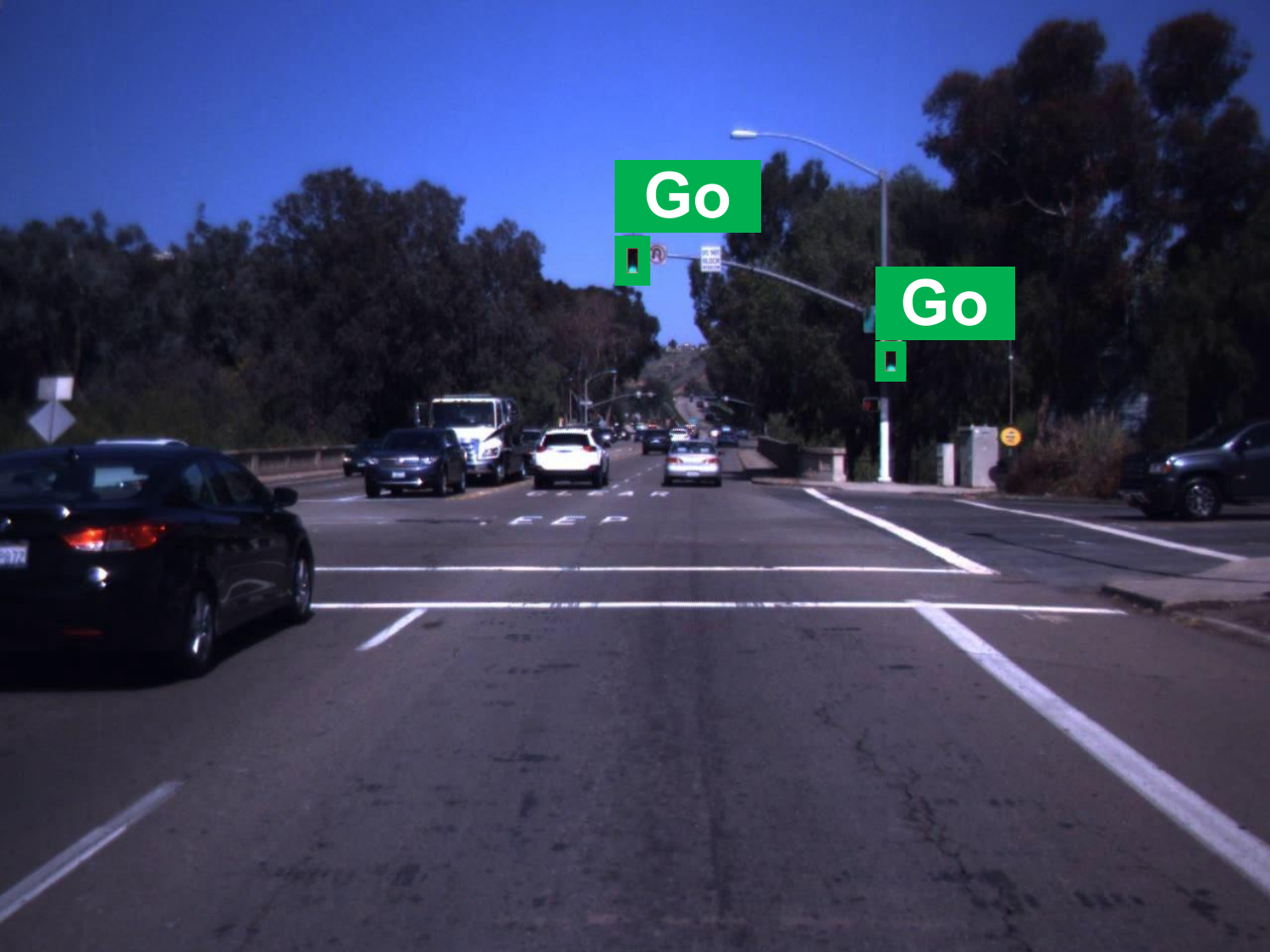}\label{fig:rq1:errors:a}}
   \subfigure[LF]{
      \includegraphics[width=0.15\linewidth]{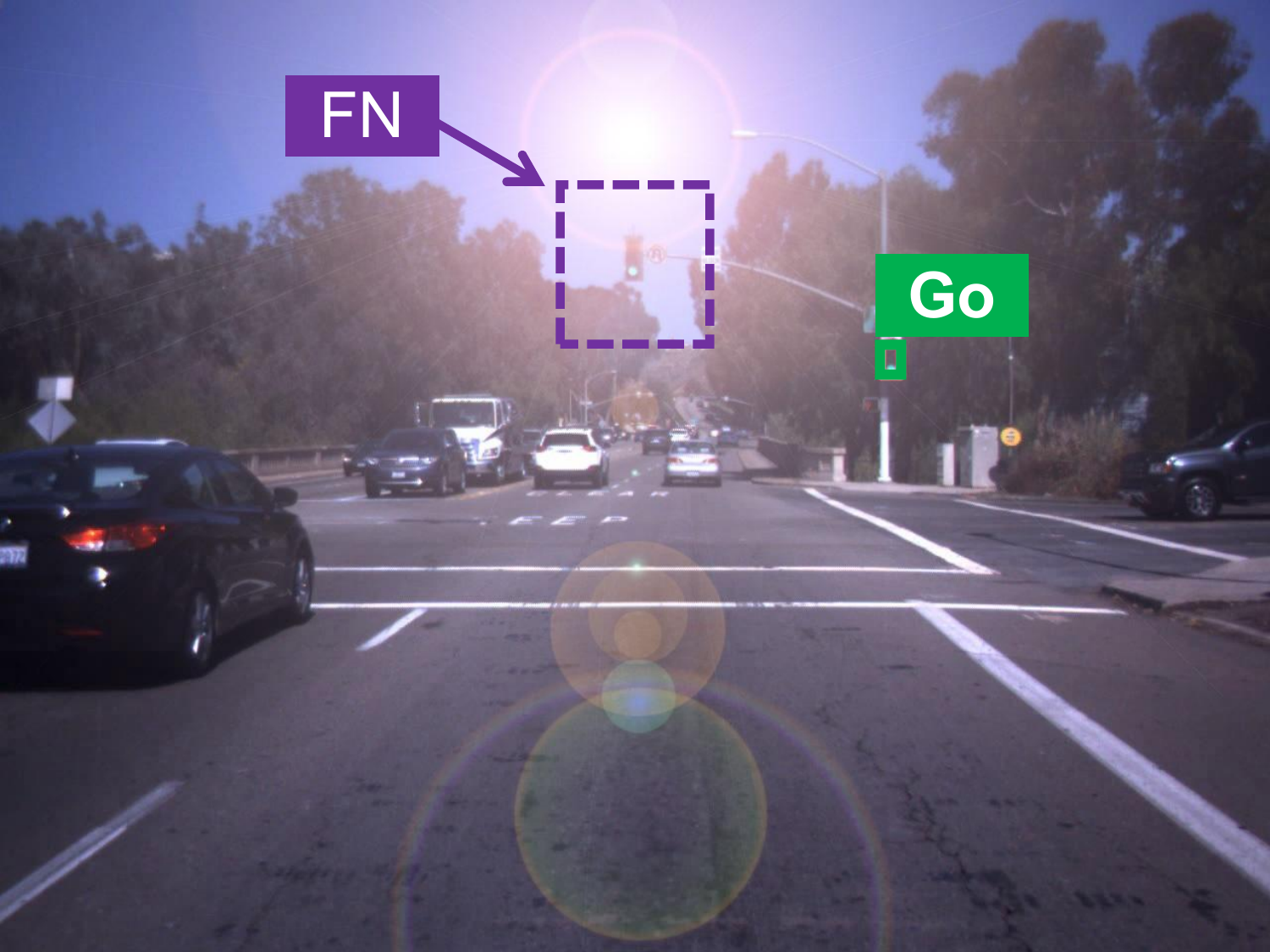}\label{fig:rq1:errors:b}}
    \subfigure[OI]{
        \includegraphics[width=0.15\linewidth]{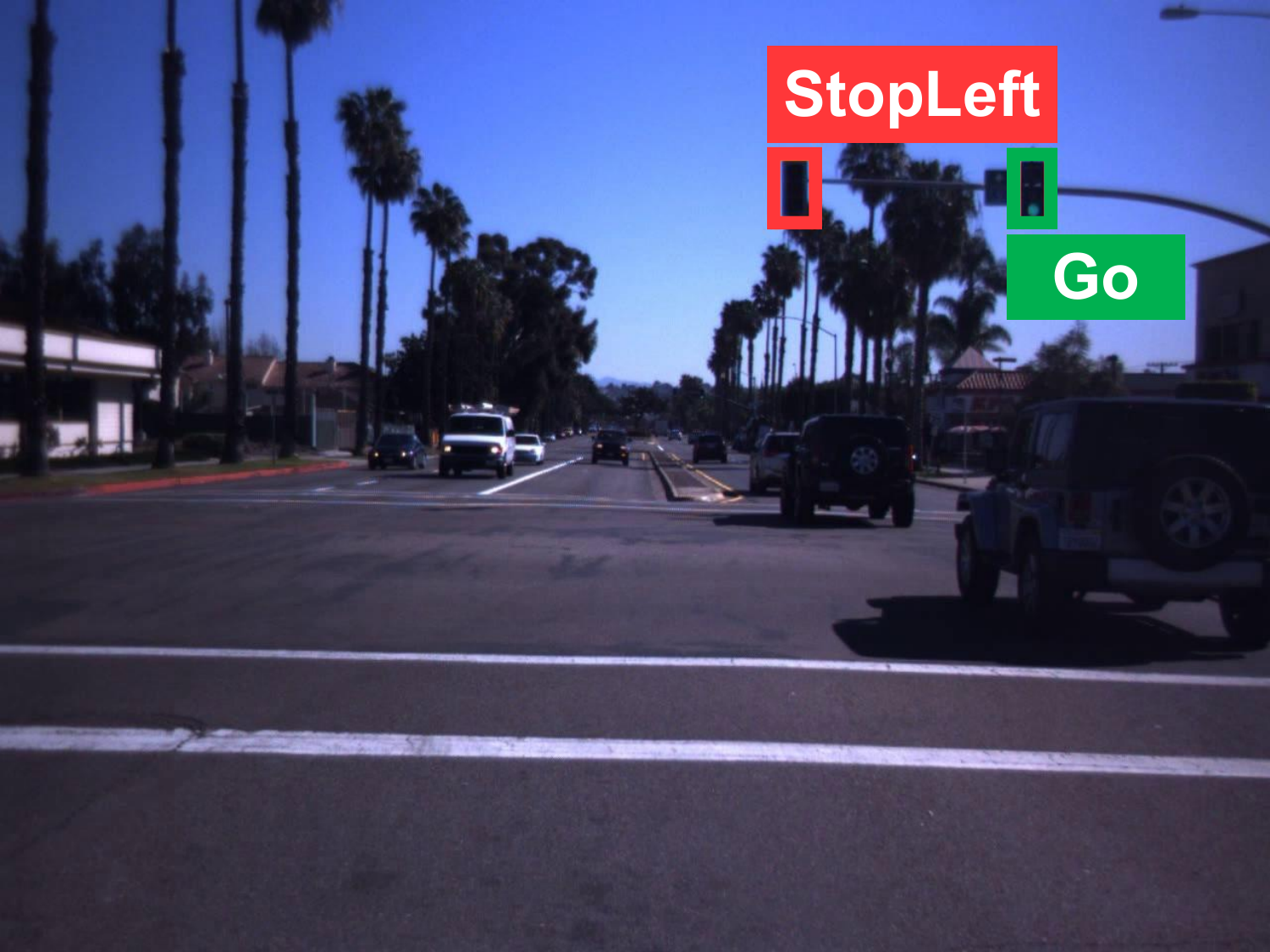}\label{fig:rq1:errors:c}}
    \subfigure[SW]{
        \includegraphics[width=0.15\linewidth]{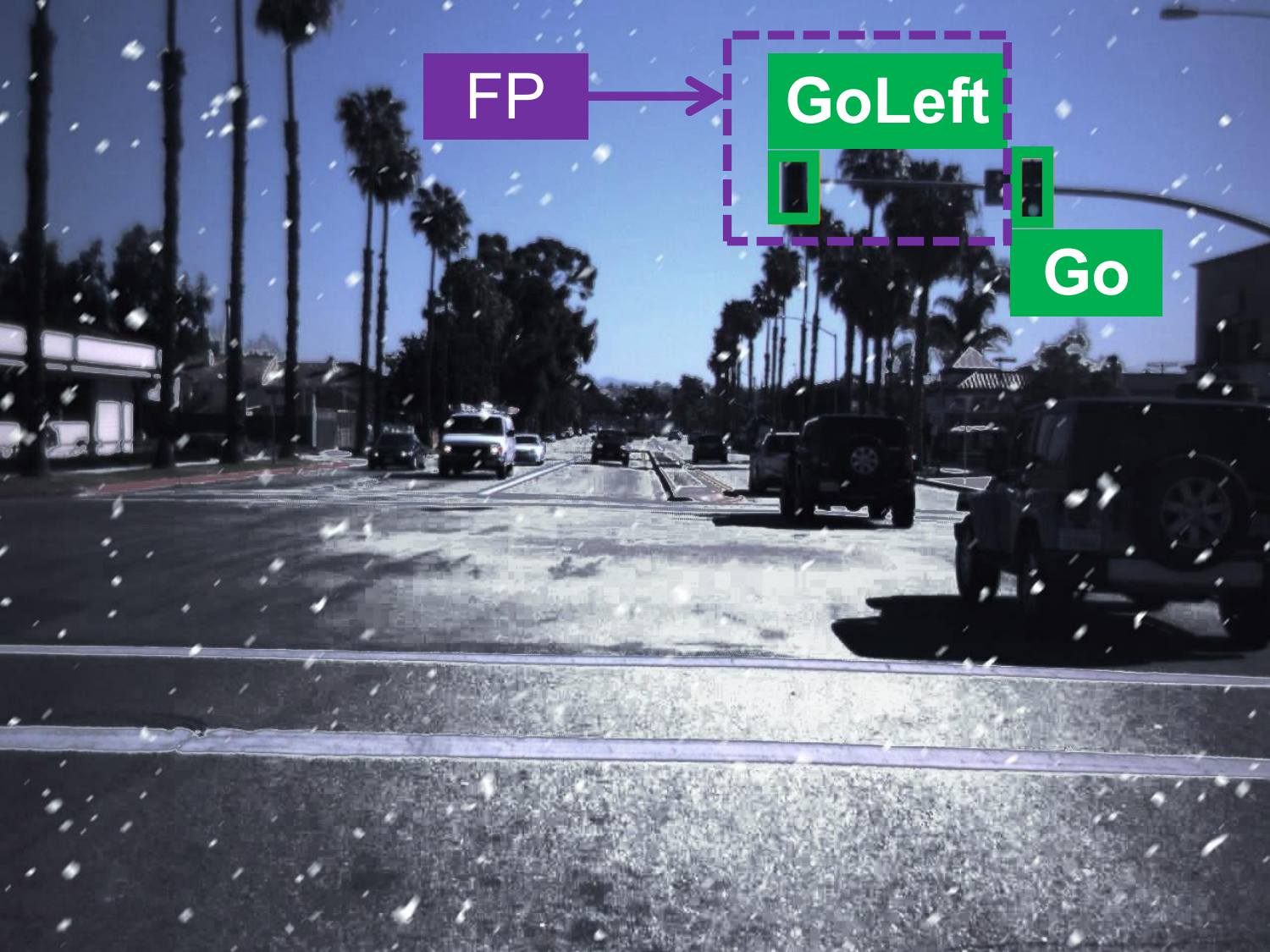}\label{fig:rq1:errors:d}}
    \subfigure[OI]{
        \includegraphics[width=0.15\linewidth]{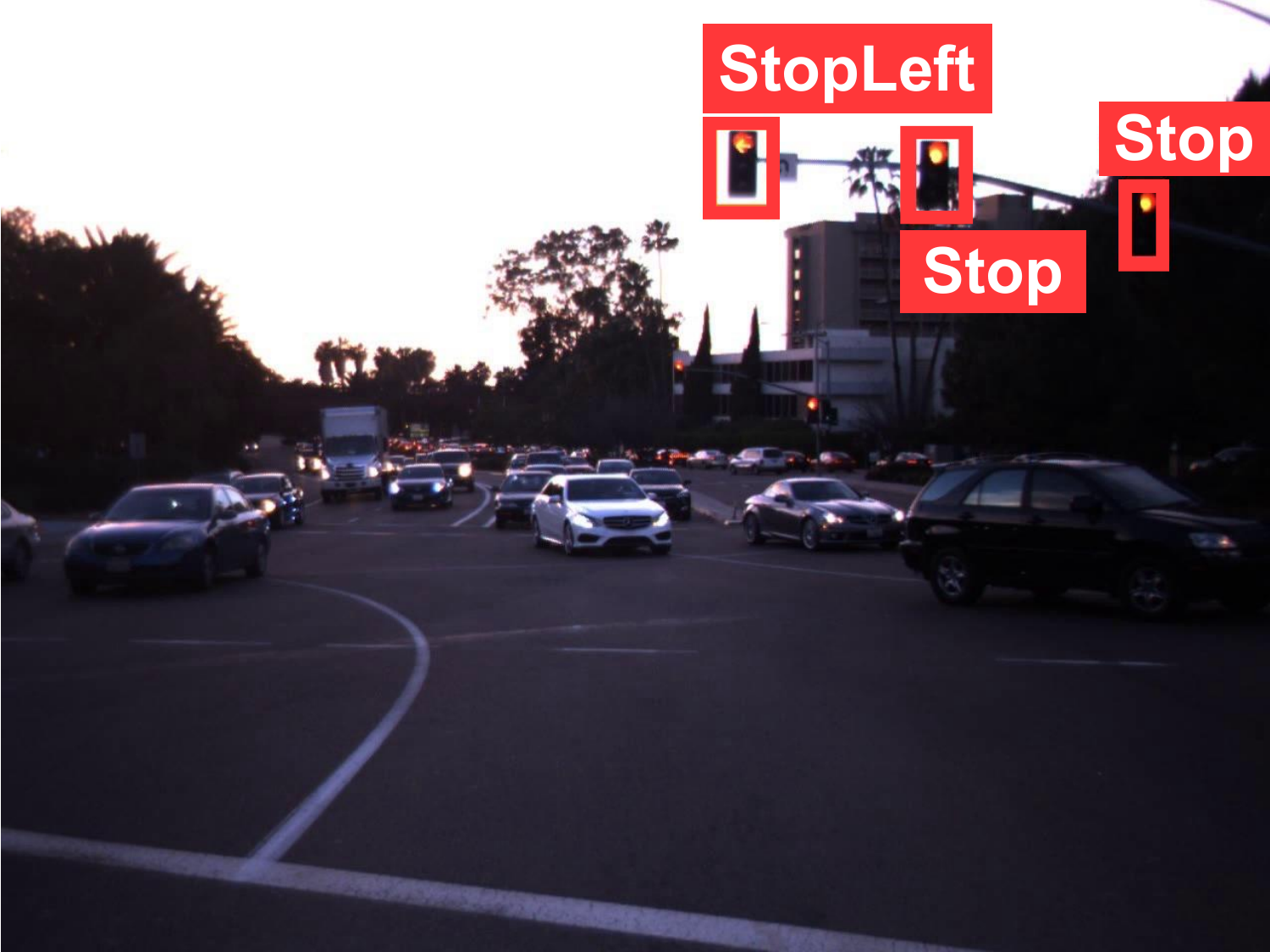}\label{fig:rq1:errors:e}}
    \subfigure[RN]{
        \includegraphics[width=0.15\linewidth]{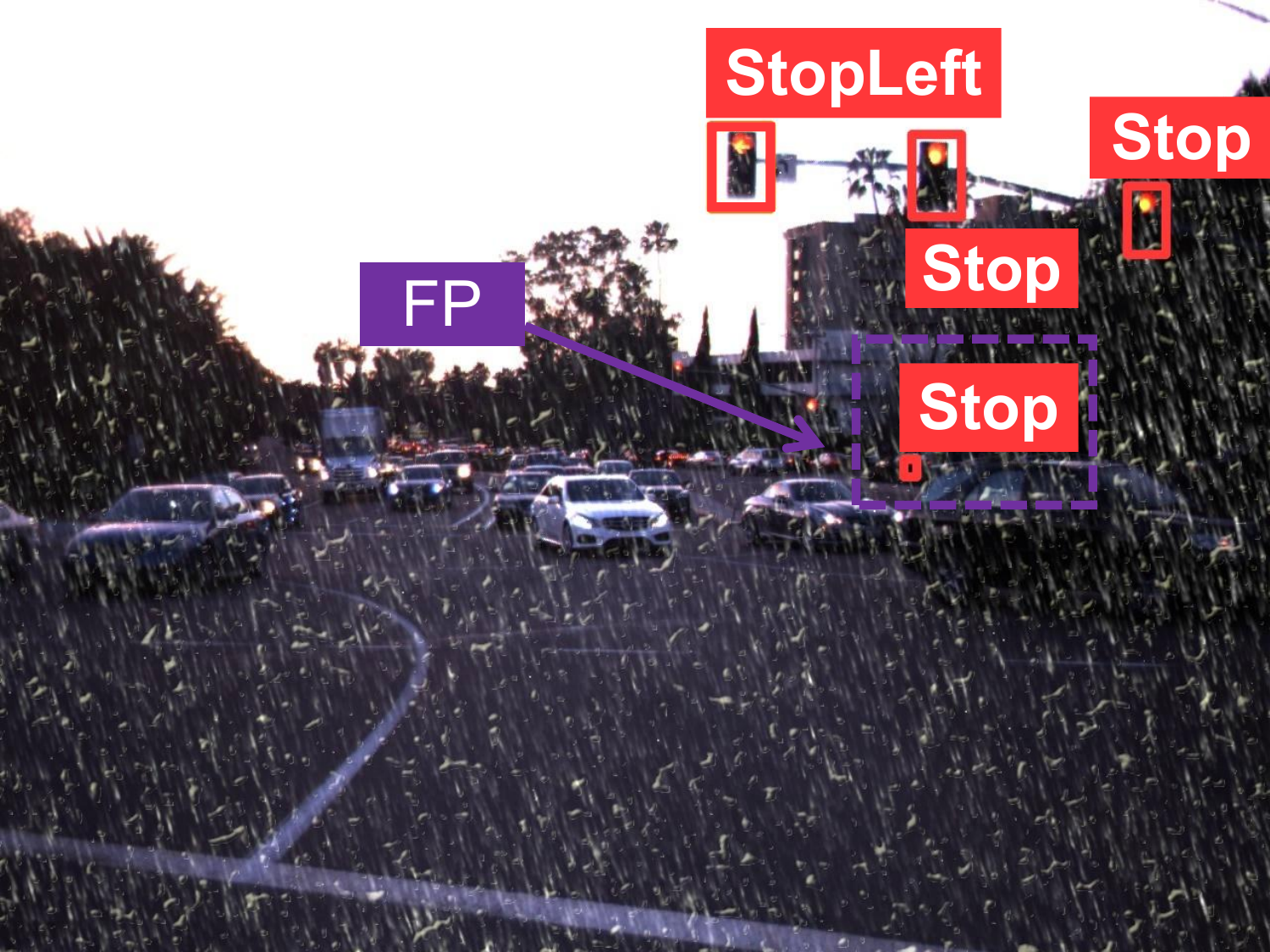}\label{fig:rq1:errors:f}}
    \subfigure[OI]{
        \includegraphics[width=0.15\linewidth]{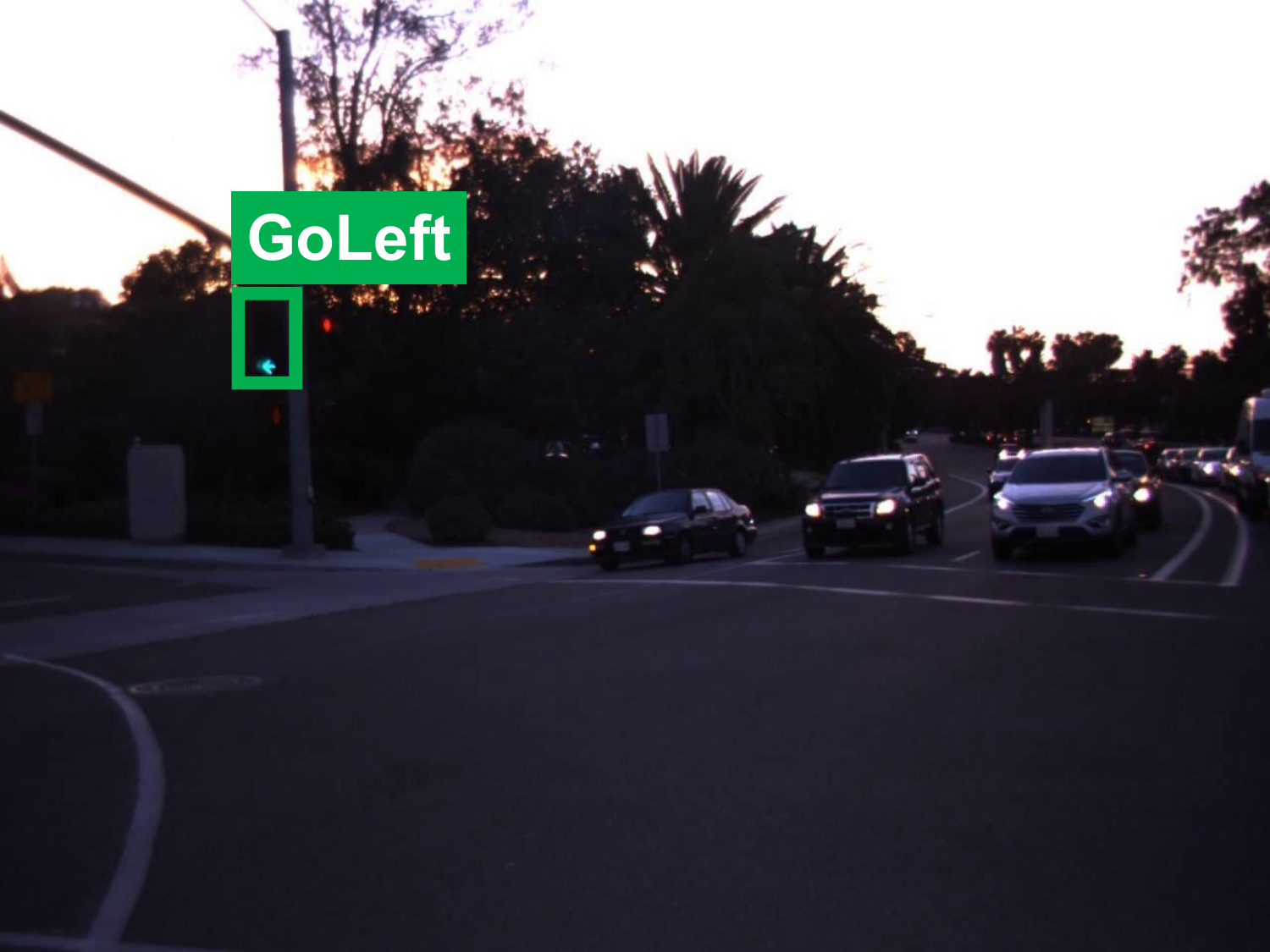}\label{fig:rq1:errors:g}}
    \subfigure[CC]{
\includegraphics[width=0.15\linewidth]{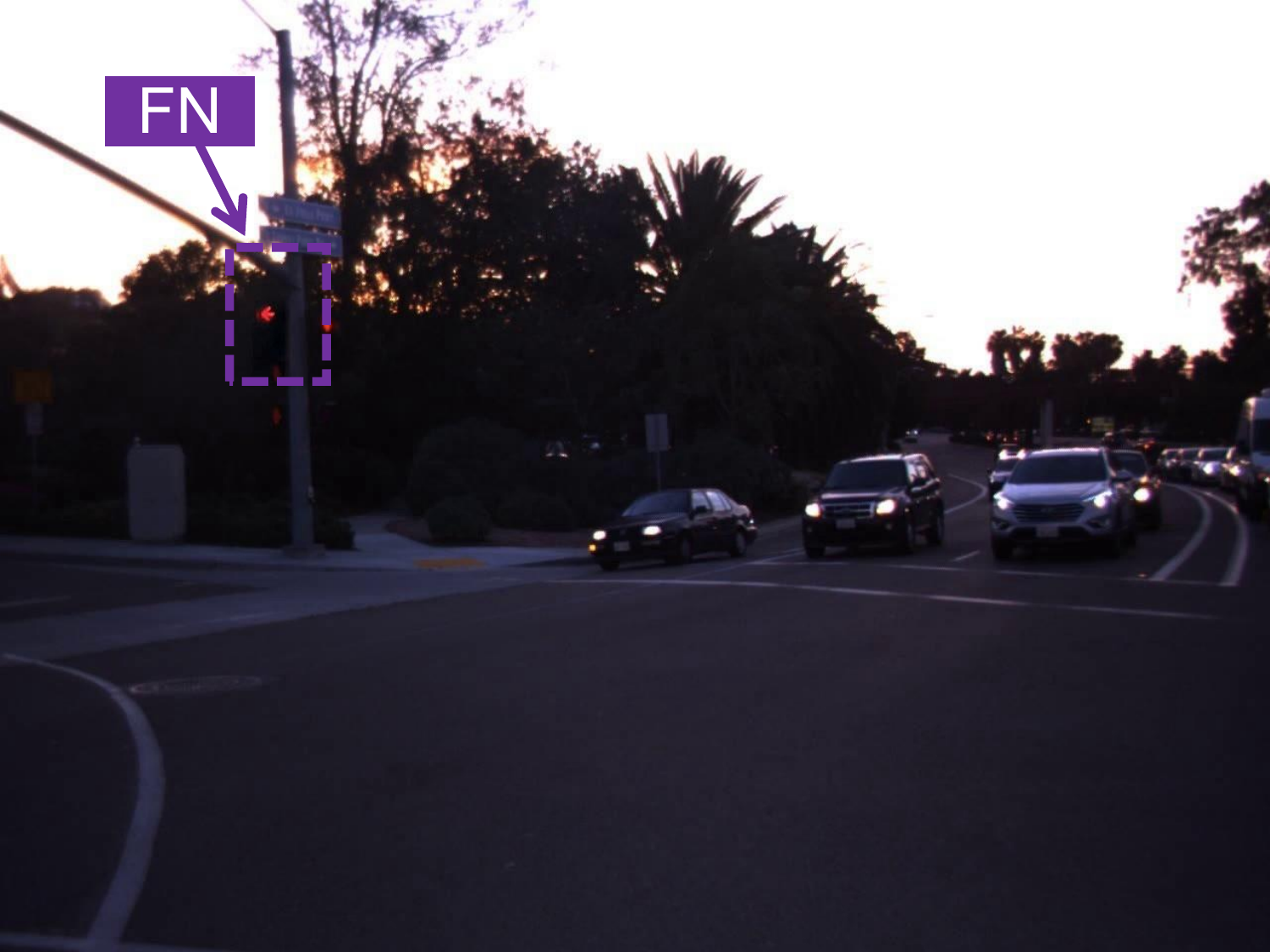}\label{fig:rq1:errors:h}}
    \subfigure[OI]{
  \includegraphics[width=0.15\linewidth]{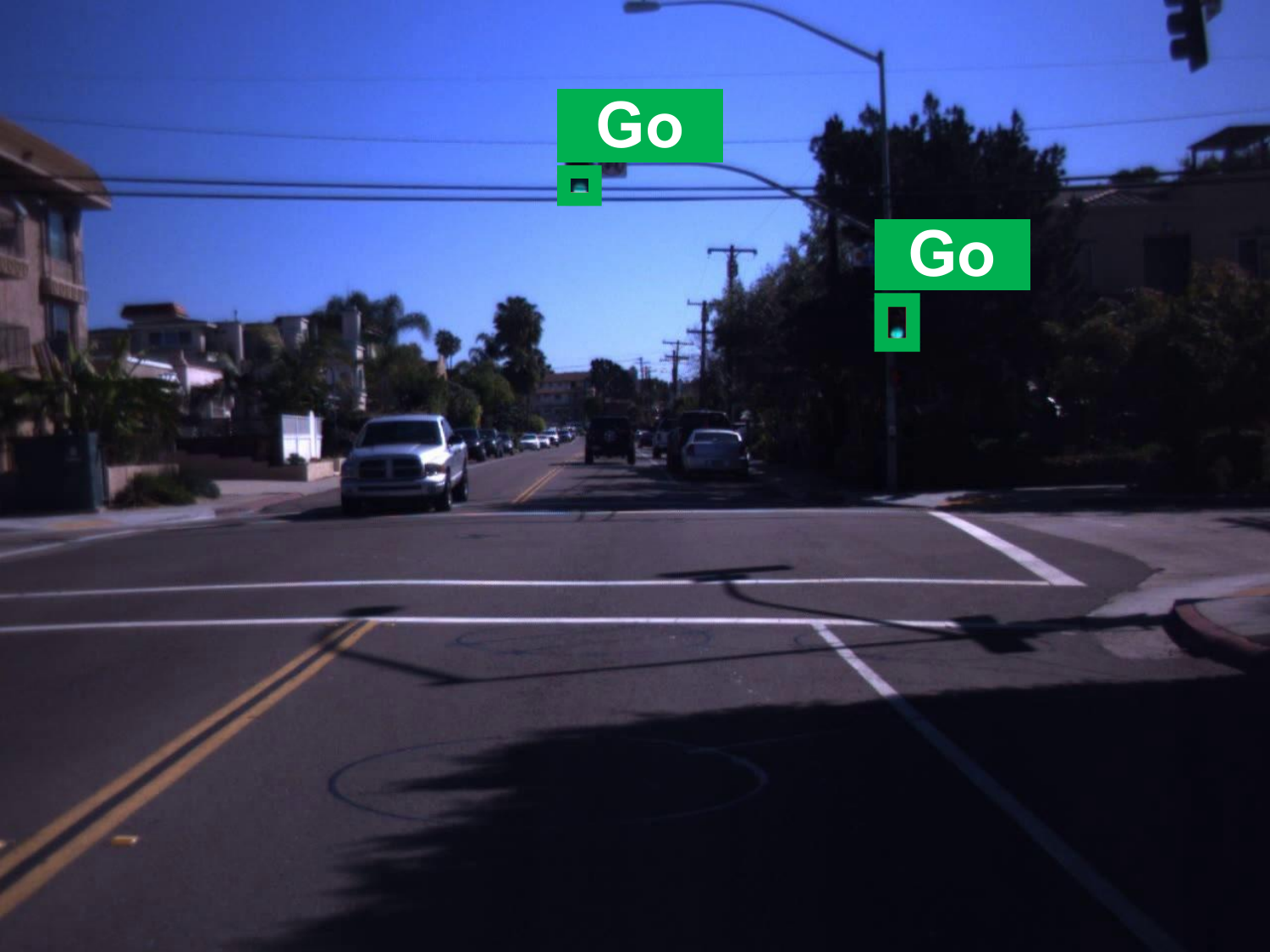}\label{fig:rq1:errors:i}}
    \subfigure[CC]{
     \includegraphics[width=0.15\linewidth]{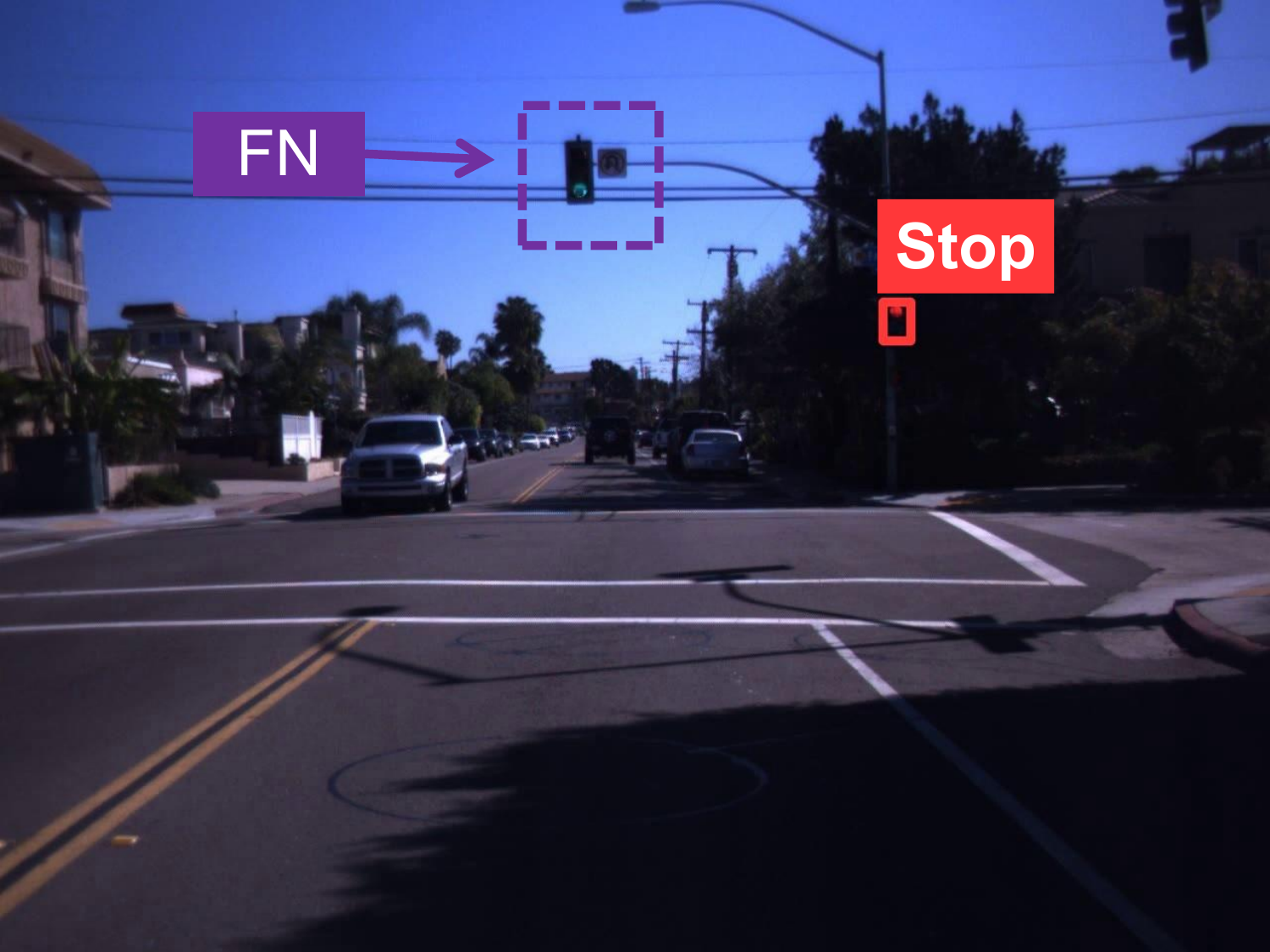}\label{fig:rq1:errors:j}}
    \subfigure[OI]{
  \includegraphics[width=0.15\linewidth]{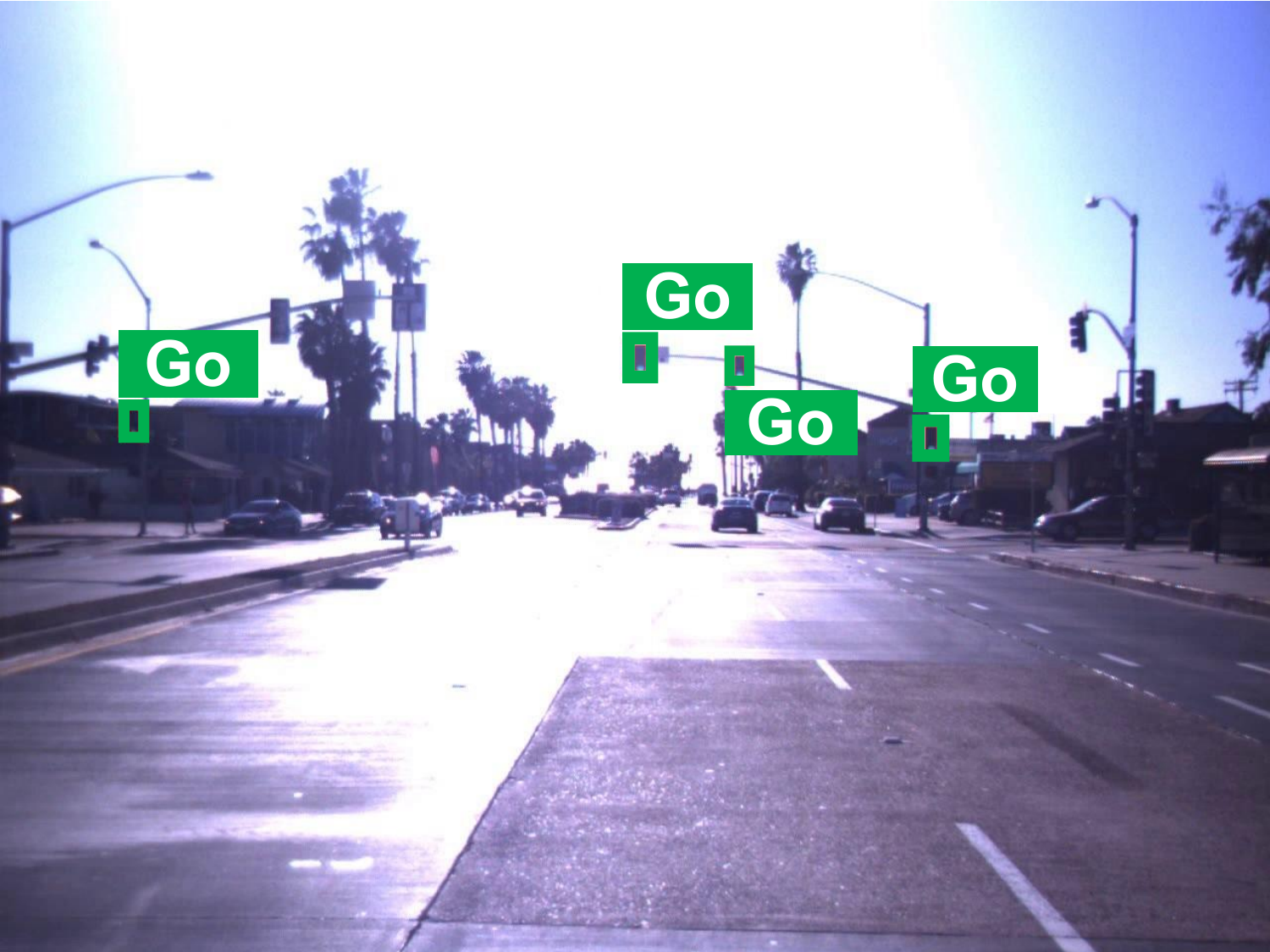}\label{fig:rq1:errors:k}}
 \subfigure[RT]{
  \includegraphics[width=0.15\linewidth]{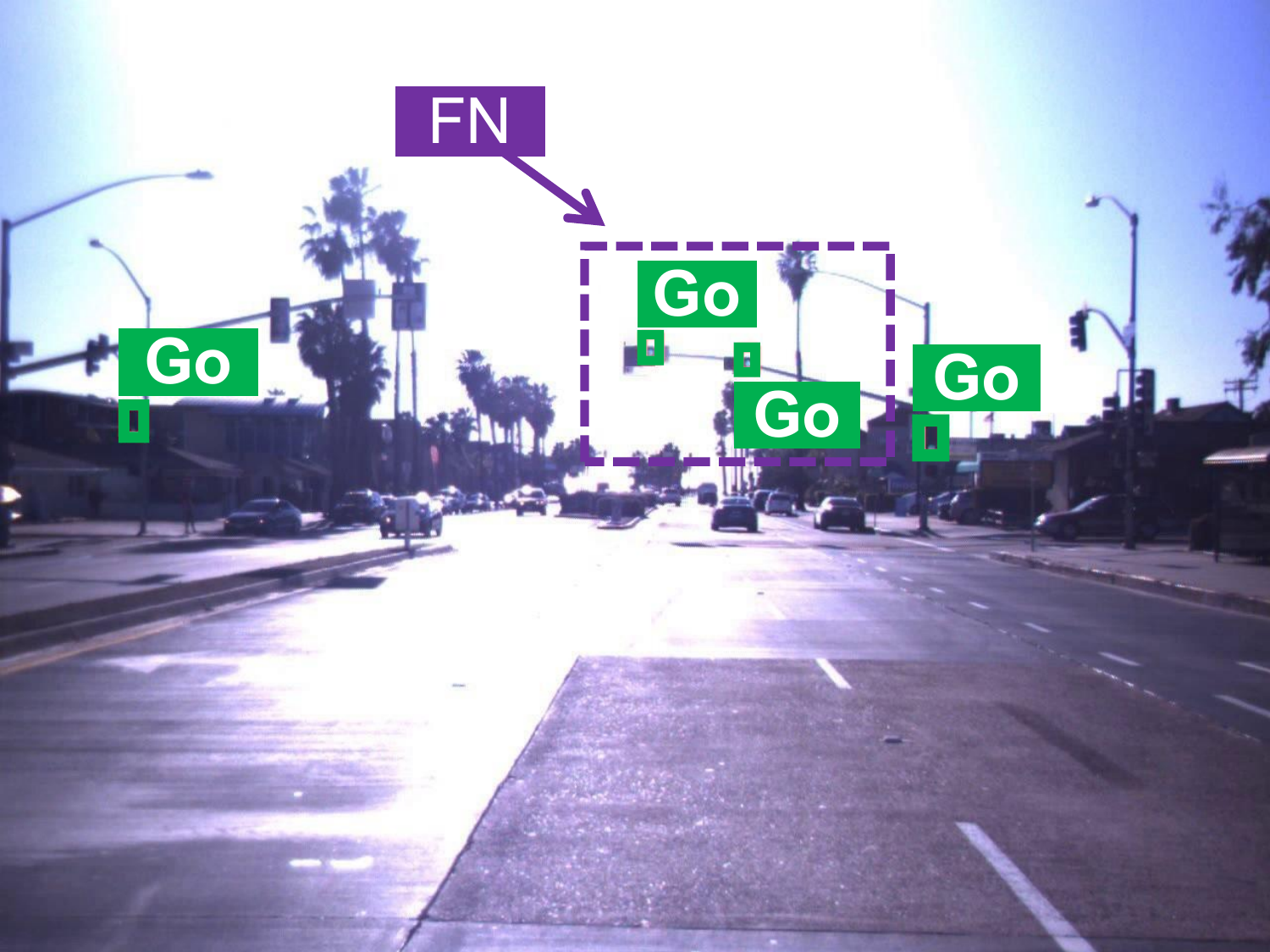}\label{fig:rq1:errors:l}}
    % \vspace{-2pt}
    \caption{Samples of Traffic Light Images Revealing Erroneous Behaviors Found by \tool}\label{fig:rq1:errors}
\end{figure*}

\textbf{mAP Drops of Original Models.} Fig. \ref{fig:rq1} reports the mAP~of the original models on the original and augmented testing datasets.~The~mAP of all the four models on the augmented testing datasets decreases significantly across all the twelve transformations, when compared with that on the original testing datasets. Such mAP drops indicate that the original models are not robust to augmented images.

% For example, in Fig. \ref{fig:rq1_1}, the original \textit{YOLOv5} model achieves a mAP of about \todo{0.35} using \texttt{RN+} as testing dataset, declining significantly with a mAP of \todo{RT+} compared to \texttt{O} as testing dataset. 

With respect to different models, \textit{YOLOv5}, \textit{YOLOX}, \textit{Faster R-CNN} and \textit{SSD} respectively suffer an average mAP drop of \todo{31.4\%}~(from~\todo{0.615} on the original testing dataset to \todo{0.422} on the augmented testing dataset), \todo{31.7\%} (from \todo{0.624} to \todo{0.426}), \todo{53.9\%} (from \todo{0.459} to \todo{0.212}),~and \todo{46.7\%} (from \todo{0.315} to \todo{0.168}) on \textit{LISA} across all transformations.~Similarly, they respectively suffer an average mAP drop of \todo{42.4\%} (from \todo{0.283} to \todo{0.163}), \todo{29.1\%} (from \todo{0.220} to \todo{0.156}), \todo{48.0\%} (from \todo{0.155}~to~\todo{0.081}), and \todo{35.4\%} (from \todo{0.057} to \todo{0.037}) on \textit{Bosch}. Moreover, \textit{YOLOv5}, \textit{YOLOX}, \textit{Faster R-CNN} and \textit{SSD} respectively suffer a mAP drop between~\todo{0.031} (for \todo{UE} on \todo{\textit{Bosch}}) and \todo{0.443} (for \todo{RT} on \todo{\textit{LISA}}), \todo{0.013} (for \todo{FG} on \todo{\textit{Bosch}}) and \todo{0.464} (for \todo{RT} on \todo{\textit{LISA}}), \todo{0.026} (for \todo{UE} on \todo{\textit{Bosch}}) and \todo{0.388} (for~\todo{FG} on \todo{\textit{LISA}}), and \todo{0.005} (for \todo{UE} on \todo{\textit{Bosch}}) and \todo{0.212} (for \todo{RT} on \todo{\textit{LISA}}).

% \todo{\textit{Faster R-CNN}} models has a maximal decrease of average mAP with \todo{0.254} in the augmented datasets of \textit{LISA} compared to the original dataset. \todo{\textit{SSD}} model has a minimal decrease of average mAP with \todo{0.152} in the augmented datasets of \textit{LISA} compared to the original dataset. 

% Similarly, in the augmented and original \textit{Bosch} datasets, the \todo{\textit{YOLOv5}} model has a maximal decrease of average mAP with \todo{0.116} and the \todo{\textit{SSD}} model has a minimal decrease of average mAP with \todo{0.021}.

With respect to different transformations, the mAP decreases by \todo{42.6\%}, \todo{41.8\%} and \todo{36.5\%} for weather, camera and traffic light transformations, respectively. The mAP decreases most significantly for \todo{FG} (an average drop of \todo{50.3\%}) among the four weather transformations, for \todo{MB} (an average drop of \todo{63.5\%}) among the three camera transformations, and for \todo{RT} (an average drop of \todo{63.9\%}) among the five traffic light transformations.

% the top error-prone transformation is \todo{rotating traffic lights (\texttt{RT+})}, which has led to a drop of average mAP to \todo{0.365} among four original models on the augmented \textit{LISA} dataset. Meanwhile, the top error-prone transformation is \todo{motion blur (\texttt{MB+})} in the augmented \textit{Bosch} dataset, which has led to an average mAP drop to \todo{0.119}.

\textbf{Detected Erroneous Behaviors.} We further analyze the erroneous behaviors detected by \tool that violate metamorphic~relations, and summarize the false positives and false negatives reported by the original models on augmented testing datasets. \ly{Fig. \ref{fig:rq1:errors} shows samples of traffic light images revealing erroneous behaviors, where the green and red boxes are generated by the traffic light detection models and the purple boxes are annotated by us.} All transformations from the three families reveal erroneous behaviors. On one hand, regarding weather and camera transformations, there are three kinds of erroneous behaviors. First, the original~model~fails to detect~a~traffic~light, resulting in a false negative, as illustrated~by Fig. \ref{fig:rq1:errors:a} and \ref{fig:rq1:errors:b}. Second, the original model predicates a false label for a traffic light, resulting in a false positive, as illustrated by Fig. \ref{fig:rq1:errors:c} and \ref{fig:rq1:errors:d}. Third, the original model recognizes~a~car's~back~light as a traffic light, resulting in a false positive, as shown~by~Fig.~\ref{fig:rq1:errors:e} and \ref{fig:rq1:errors:f}. On the other hand, regarding traffic light transformations, there are also three kinds of erroneous behaviors. First, the original model fails to detect a transformed traffic light, causing a false negative, as illustrated by Fig. \ref{fig:rq1:errors:g} and \ref{fig:rq1:errors:h}. Second,~the~original model detects a transformed traffic light successfully, but fails to detect the unchanged traffic light or generates a wrong label,~leading to a false negative or false positive, as shown by Fig. \ref{fig:rq1:errors:i} and \ref{fig:rq1:errors:j}. Third, the original model detects a transformed traffic light,~but finds a bounding box that is far too different from the ground~truth, resulting in a false negative, as illustrated~by~Fig.~\ref{fig:rq1:errors:k}~and~\ref{fig:rq1:errors:l}.

\begin{tcolorbox}[size=title, opacityfill=0.15]
\textit{\textbf{Summary.}} All the four models suffer a decrease~in~mAP~on the augmented testing datasets across all the twelve transformations on the two datasets. On average,~the~mAP of each model decreases by \todo{39.8\%}. Moreover, these~models are more prone to erroneous behaviors for \todo{weather} and \todo{camera} transformations than for \todo{traffic light} transformations. The average mAP drop ranges from \todo{17.1\%} for \todo{SC} to \todo{63.9\%}~for~\todo{RT}. Therefore, \tool is effective in detecting erroneous behaviors for all models.
\end{tcolorbox}

\subsection{Retraining Evaluation (RQ2)}

% To observe the accuracy of the retrained models, we use both the original testing datasets and the augmented testing datasets. The result is respectively illustrated in Fig. \ref{fig:rq2} and Fig. \ref{fig:rq3}.

\begin{figure*}[!t]
    \centering
    % \subfigcapskip=-4pt
    \subfigure[Results on the \textit{LISA} Dataset]{
		\includegraphics[width=0.48\linewidth]{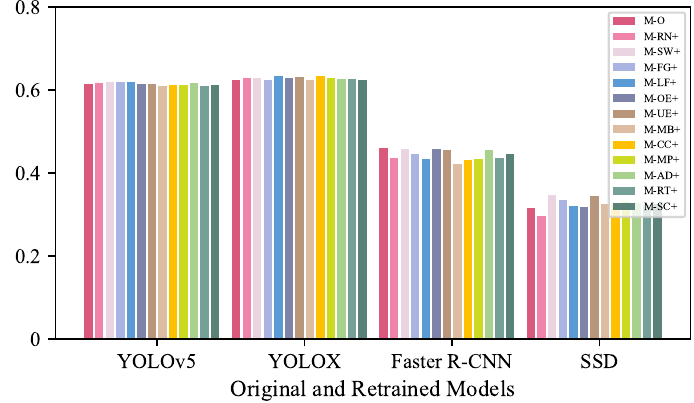}\label{fig:rq2_1}}
    \subfigure[Results on the \textit{Bosch} Dataset]{
            \includegraphics[width=0.48\linewidth]{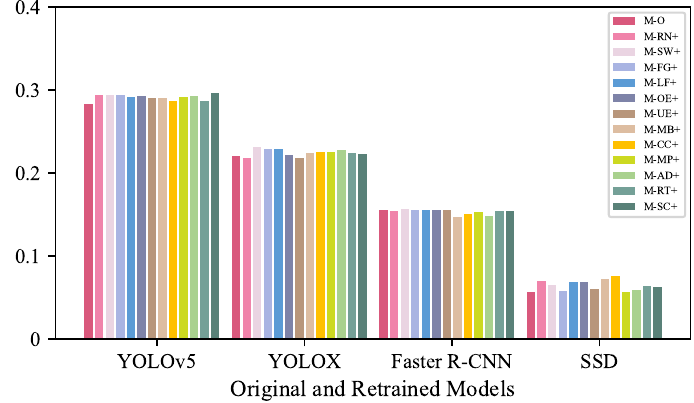}\label{fig:rq2_2}}
    \vspace{-7pt}
    \caption{mAP Comparison between Original and Retrained Models on the Original Testing Datasets}\label{fig:rq2}
\end{figure*}

\textbf{mAP Changes on the Original Datasets.} Fig.~\ref{fig:rq2} shows the mAP~of the original and retrained models on the original testing datasets. Overall, the retrained models achieve a similar mAP as the original models on the original testing datasets. 

Specifically, with respect to retrained models, the average mAPs of retrained \textit{YOLOv5}, \textit{YOLOX} and \textit{SSD} models across all transformations reach \todo{0.615}, \todo{0.628} and \todo{0.326} on the~original \textit{LISA} testing dataset (see Fig.~\ref{fig:rq2_1}), which are slightly higher~by \todo{1.4\%} than~those~of the original models. Only the average~mAP of retrained \textit{Faster R-CNN} models become lower;~i.e.,~the~average mAP is \todo{0.443} (only \todo{3.6\%} lower than the original model). Similarly, the average mAPs of retrained \textit{YOLOv5}, \textit{YOLOX} and \textit{SSD} models are \todo{0.292}, \todo{0.225} and \todo{0.065} on the original \textit{Bosch} testing dataset (see Fig.~\ref{fig:rq2_2}), which are slightly higher by \todo{6.4\%} than those of the original models. Only the average mAP of retrained \textit{Faster R-CNN} models is lower; i.e., the average mAP is \todo{0.153} (only \todo{1.3\%} lower than the original model). 

With respect to transformations, the average mAPs~of~retrained models slightly increase for all the 12 transformations compared with the original models. \todo{SW} causes a largest increase of average mAP by \todo{4.4\%}. \todo{MP} causes a smallest increase of average mAP~by~\todo{0.3\%}.

\begin{figure*}[!t]
    \centering
    % \vspace{-5pt}
    % \subfigbottomskip=-1pt 
    % \subfigcapskip=-2pt
    % \footnotesize
    
    \subfigure[\scriptsize \textit{YOLOv5} on \textit{LISA}]{
		\includegraphics[width=0.23\linewidth]{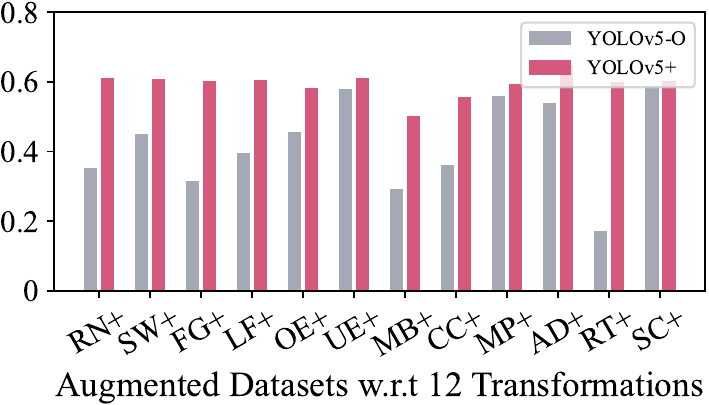}}\label{fig:rq3_1}
    \subfigure[\scriptsize \textit{YOLOX} on \textit{LISA}]{
            \includegraphics[width=0.23\linewidth]{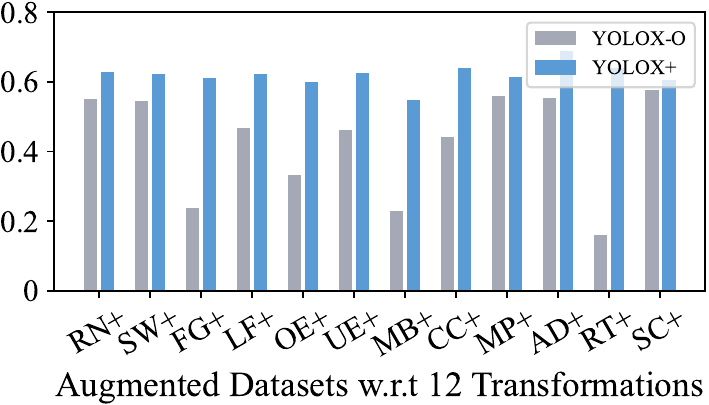}}\label{fig:rq3_2}
    \subfigure[\scriptsize \textit{Faster R-CNN} on \textit{LISA}]{
            \includegraphics[width=0.23\linewidth]{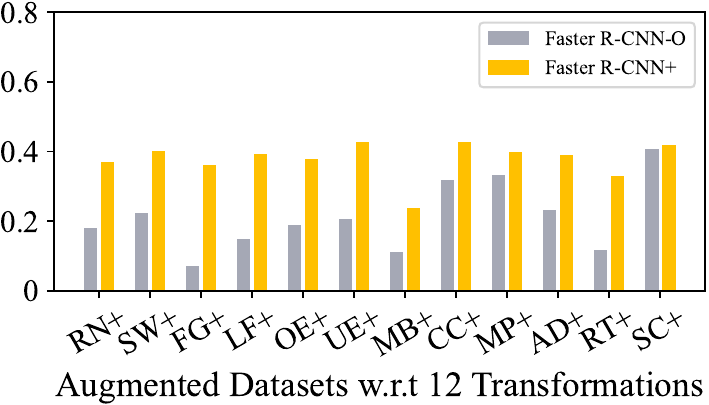}}\label{fig:rq3_3}
    \subfigure[\scriptsize \textit{SSD} on \textit{LISA}]{
            \includegraphics[width=0.23\linewidth]{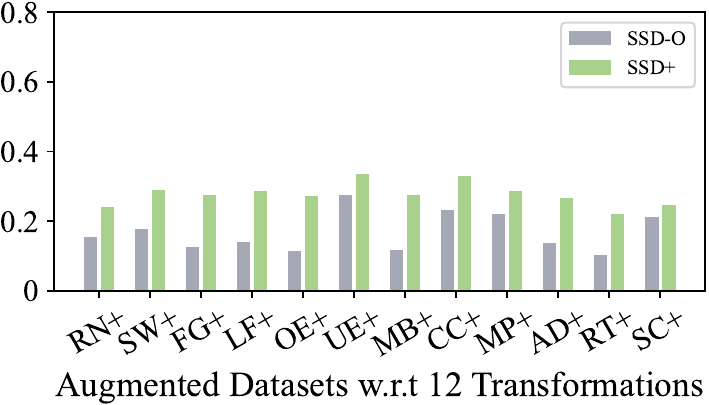}}\label{fig:rq3_4}
    \subfigure[\scriptsize \textit{YOLOv5} on \textit{Bosch}]{
        \includegraphics[width=0.23\linewidth]{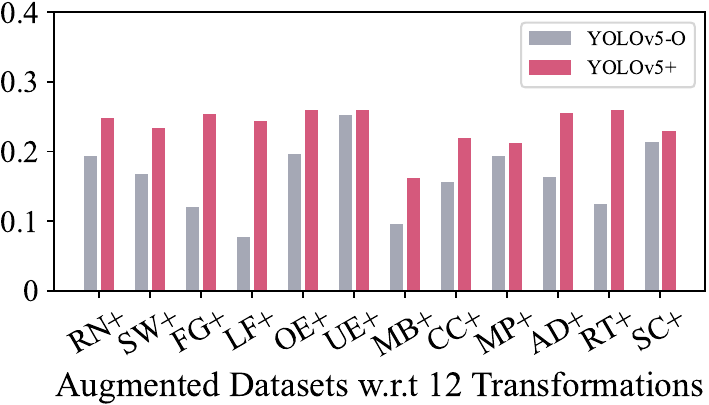}}\label{fig:rq3_5}
    \subfigure[\scriptsize \textit{YOLOX} on \textit{Bosch}]{
            \includegraphics[width=0.23\linewidth]{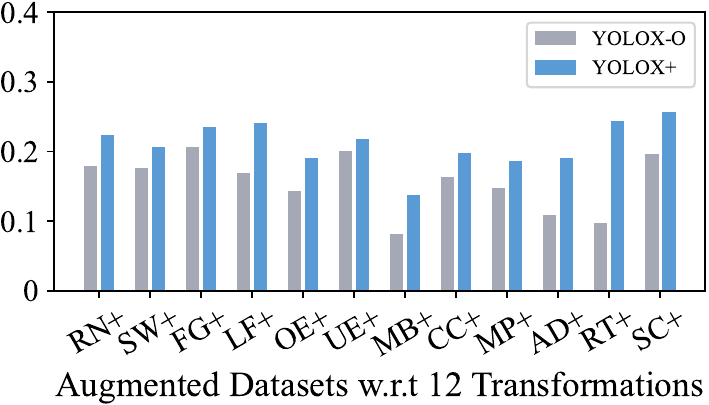}}\label{fig:rq3_6}
    \subfigure[\scriptsize \textit{Faster R-CNN} on \textit{Bosch}]{
            \includegraphics[width=0.23\linewidth]{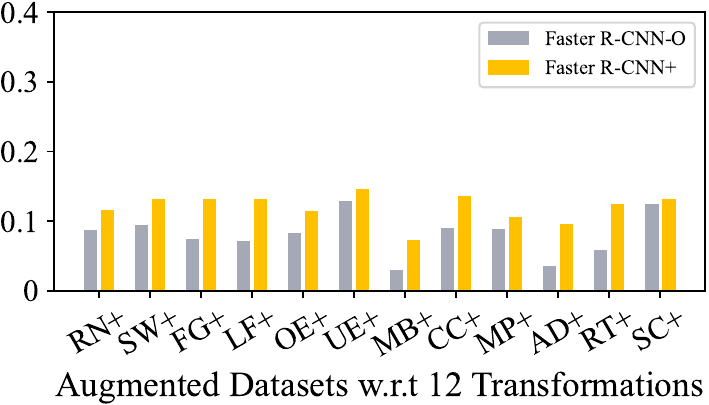}}\label{fig:rq3_7}
    \subfigure[\scriptsize \textit{SSD} on \textit{Bosch}]{
            \includegraphics[width=0.23\linewidth]{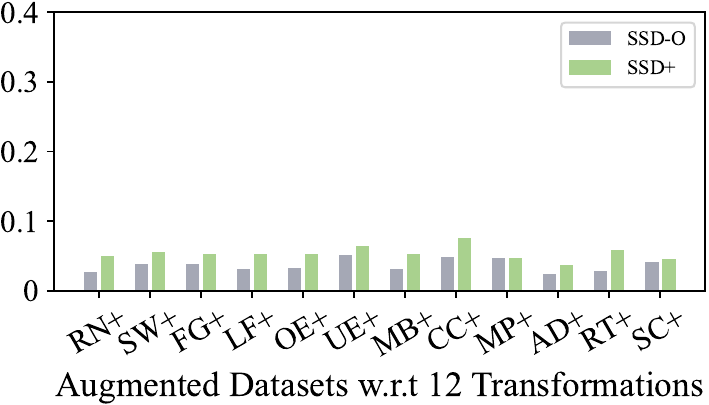}}\label{fig:rq3_8}
    % \vspace{-2pt}
    \caption{mAP Comparison between Original and Retrained Models on the Augmented Testing Datasets}\label{fig:rq3}
\end{figure*}

\textbf{mAP Changes on the Augmented Datasets.} Fig.~\ref{fig:rq3}~gives~the mAP~of the original and retrained models on the augmented testing datasets. All retrained models~have~a~significantly~higher mAP than the original models on augmented testing datasets. 

% the mAPs of the original models decrease significantly on augmented datasets while the retrained models obtains a similar result compared with the ones on the original datasets. 

Specifically, with respect to retrained models, the retrained \textit{Faster R-CNN} models obtain the highest improvement~in~the average~mAP among the four models, i.e., an average mAP improvement~of~\todo{115.5\%} and~\todo{65.5\%} on the augmented \textit{LISA} and \textit{Bosch} testing datasets. Meanwhile,~the retrained \textit{YOLOv5}, \textit{YOLOX} and \textit{SSD} models respectively~have an average mAP improvement of \todo{57.1\%} and~\todo{60.1\%}, \todo{71.5\%} and~\todo{42.8\%}, and \todo{76.8\%} and~\todo{51.0\%} on the augmented \textit{LISA} and \textit{Bosch} testing~datasets.

With respect to transformations, \todo{RT} contributes the most in improving the original models. It improves the average mAP by \todo{210.5\%} and \todo{118.7\%} on the augmented \textit{LISA} and \textit{Bosch} testing datasets,~followed by \todo{FG} with an average~mAP~improvement of \todo{193.5\%} and \todo{59.6\%}. \todo{SC} contributes the least improvement, i.e., an average~mAP~improvement of \todo{6.4\%} and \todo{13.6\%} on the augmented \textit{LISA} and \textit{Bosch} testing datasets.

%\begin{figure}[!t]
%    \centering
%    \subfigcapskip=-3pt
%    \subfigure[Time of Transformations]{
%		\includegraphics[width=0.7\linewidth]{fig/rq4_1.pdf}\label{fig:rq4_1}}
%    \subfigure[Time of Model Retraining]{
%        \includegraphics[width=0.7\linewidth]{fig/rq4_2.pdf}\label{fig:rq4_2}}
%    \vspace{-15pt}
%    \caption{Time Overhead of \tool}\label{fig:rq4}
%\end{figure}
\begin{figure}[!t]
    \centering
    \includegraphics[width=0.8\linewidth]{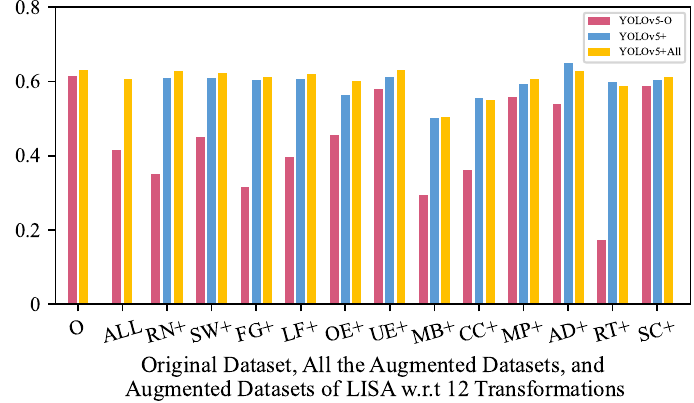}
    \vspace{-8pt}
    \caption{mAP Comparison between Original Model, Models Retrained~by~Augmented Datasets Separately, and the Model Retrained by Putting All~Augmented Datasets Together on Original, Augmented and All Augmented Testing Datasets}\label{fig:rq2-all}
\end{figure}

\textbf{mAP Changes for the Model Retrained by Including~All Augmented Training Datasets Together.} Fig.~\ref{fig:rq2-all} shows the result of mAP comparison among original models (the red~bars), retrained models using the augmented training dataset generated by each of the 12 transformations separately (the blue bars),~and retrained model using all the augmented training~datasets~generated by the 12 transformations together (the yellow bars) on the original testing datasets, augmented testing datasets w.r.t 12 transformations respectively, and all the augmented testing datasets together. We report the result of \textit{YOLOv5}~on \textit{LISA}. The remaining results are similar, and are available at \url{https://tigaug.github.io/}.

First, on the original testing dataset (i.e., \texttt{O}), the retrained model using all the augmented training datasets together has a slightly higher mAP than the~original model. Second, on all the augmented testing datasets (i.e., \texttt{ALL}), the mAP~of~the retrained model using all the augmented training datasets~together reaches \todo{0.608}, which is significantly higher by \todo{46.2\%} than~that~of~the original model.~Third, on the augmented testing dataset~generated by each~of the 12 transformations~(i.e.,~\texttt{$\tau$+}),~the~retrained models achieve~a huge improvement. Particularly, the retrained model~using~all the augmented training datasets~together~mostly has a slightly higher mAP than the retrained models~using one type of augmented training dataset.

% except \todo{AD} and \todo{RT} (only \todo{3\%} and \todo{2\%} lower respectively ).

% Hence, the retrained model using all transformations performs better than models retrained by one type of transformations. The retrained model using all transformations seems more robust.

\begin{tcolorbox}[size=title, opacityfill=0.15]
\textit{\textbf{Summary.}} Retrained models achieve similar mAP as original models on the original testing datasets, but have significantly higher mAP (i.e., \todo{67.5\% higher}) than original models on the augmented testing datasets. \todo{RT} contributes the most~in~improving performance of original models. The retrained model using all transformations performs better than retrained models using one type of transformations. Thus, \tool~is~effective~in~improving performance by feeding augmented~data~for~retraining.
\end{tcolorbox}

% !TeX root = ../main.tex

\subsection{Efficiency Evaluation (RQ3)} 

We first measure the time cost of transformations. %The result~is~reported in Fig. \ref{fig:rq4_1}. 
Generally, \tool takes less than 10 hours~for~\todo{22} of the 24~transformation tasks except for two, i.e., CC on \textit{LISA} and SC~on~\textit{LISA}, each of which takes around 30 hours.~In~total,~\tool~consumes \todo{137} hours to generate 24 augmented datasets.~On~average, it takes \todo{0.88} seconds to synthesize an image. Extremely, it takes a maximal of \todo{3.03} seconds to synthesize an image by \todo{CC}.

Then, we further analyze the time cost of model retraining. We measure each model's average retraining time across each of the 12 transformations with respect to \textit{LISA} and \textit{Bosch} datasets. %The result~is~reported in Fig. \ref{fig:rq4_2}. 
We also measure the training~time~of~the~original~models for comparison. \textit{YOLOv5} consumes the least retraining time of less than 10 hours, while \textit{Faster R-CNN} consumes the most retraining time of more than \todo{30} hours using the augmented \textit{LISA} training datasets. This result also holds using the augmented \textit{Bosch} training datasets. Besides, retraining takes an average~of~\todo{12.35\%} more time than training the original models. In total, it takes \todo{1,464} hours to train 8 original models using the original datasets and retrain 96 models using the augmented datasets. On average, \tool takes \todo{14} hours to retrain~a~model using one type of transformation for improving performance.

When model is retrained using all transformations together, the average retraining time increases to \todo{36} hours. Notice that we provide detailed time cost of transformations and retraining at our website \url{https://tigaug.github.io/}.

\begin{figure}[!t]
    \centering
    \begin{minipage}[t]{0.48\textwidth}
    \includegraphics[width=\linewidth]{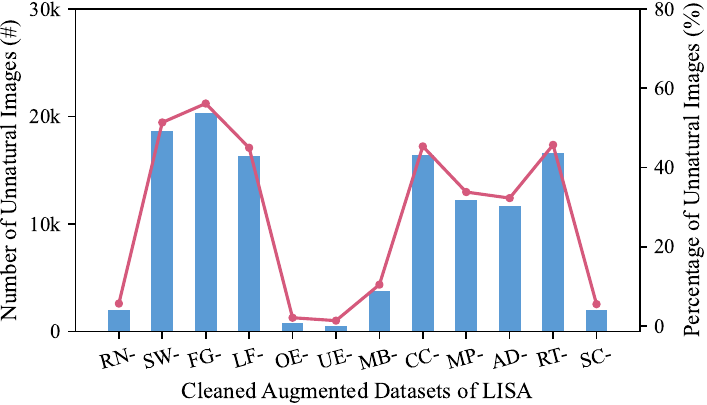}
    % \vspace{-9pt}
    \caption{Number and Percentage of Unnatural Images}\label{fig:rq5a}
    \end{minipage}
    \hfill
    \begin{minipage}[t]{0.48\textwidth}
        \includegraphics[width=0.97\linewidth]{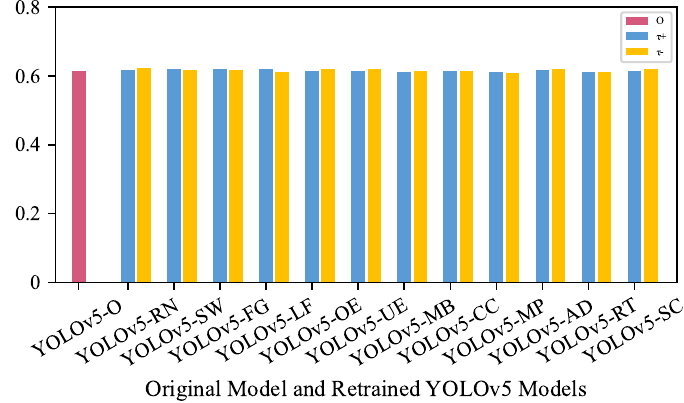}
        \caption{mAP of the Original Model and Retrained \textit{YOLOv5} Models (by~One Transformation) on the Original \textit{LISA} Testing Dataset}\label{fig:rq5b}
    \end{minipage}

\end{figure}

\begin{tcolorbox}[size=title, opacityfill=0.15]
\textit{\textbf{Summary.}} On average, \tool takes \todo{0.88} seconds~to~synthesize an image, and respectively takes \todo{14 and 36} hours~to~retrain a model by one transformation and all transformations. Given the effectiveness in detecting erroneous behaviors and improving model performance, the time cost is acceptable.
\end{tcolorbox}

% \begin{figure}[!t]
%     \centering
%        \subfigcapskip=-3pt
% %    \subfigure[\textit{YOLOv5} Models]{
% 		\includegraphics[width=0.8\linewidth]{fig/rq5_3.pdf}
% %    \subfigure[\textit{YOLOX} Models]{
% %        \includegraphics[width=0.41\linewidth]{fig/rq5_4.pdf}}
%     % \vspace{-8pt}
%     \caption{mAP of the Original Model and Retrained \textit{YOLOv5} Models (by~One Transformation) on the Original \textit{LISA} Testing Dataset}\label{fig:rq5b}
% \end{figure}

\subsection{Quality Evaluation (RQ4)}

\textbf{Number and Percentage of Unnatural Images.} We report the number and percentage of unnatural images (identified~in our manual analysis) in each augmented dataset by a transformation in Fig.~\ref{fig:rq5a}. Specifically, \tool generates the least number of unnatural images by five transformations, i.e., RN, OE, UE, MB and SC. Less~than \todo{10\%} of the augmented images by these transformations are considered as unnatural. For the remaining transformations, the percentage of unnatural images ranges from \todo{32\%} to \todo{56\%}. On average, \todo{27.9\%} of synthesized images by each transformation are considered~as~unnatural. Consequently, the number of cleaned augmented testing images is smaller than the number of original testing images by \todo{1.3\%} to \todo{56.2\%} due to the removal of unnatural images. Given~that there are \todo{6,166} original testing images, the cleaned augmented testing datasets are believed to be still sufficient.

\textbf{Impact of Unnatural Images.} Fig.~\ref{fig:rq5b} shows the result of mAP~comparisons among original models (i.e., the red bars), retrained models (using one transformation) in \textbf{RQ2} without removing unnatural data (i.e., the blue bars),~and~retrained models (using one transformation) without unnatural data (i.e., the yellow bars) on the original \textit{LISA} testing dataset. Generally, these three kinds of models achieve similar mAP, indicating that those unnatural data will not affect models' detection capability on original real-world data. 

\begin{figure*}[!t]
    \centering
   \subfigcapskip=-3pt
   \subfigbottomskip=-4pt 
%    \vspace{-5pt}
%    \subfigure[\textit{YOLOv5} Models]{
		\includegraphics[width=0.89\linewidth]{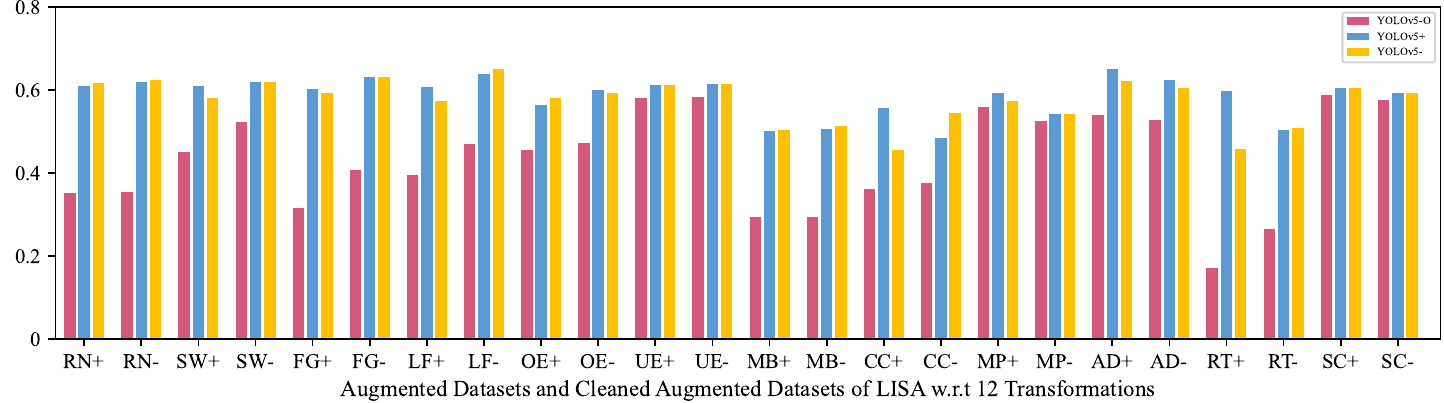}
    
%    \subfigure[\textit{YOLOX} Models]{
%        \includegraphics[width=0.89\linewidth]{fig/rq5_2.pdf}}
    % \vspace{-8pt}
    \caption{mAP of the Original Model and Retrained \textit{YOLOv5} Models (by One Transformation) on the Augmented and Cleaned Augmented \textit{LISA} Testing Dataset}\label{fig:rq5c}
\end{figure*}

Furthermore, similar to Fig.~\ref{fig:rq5b}, Fig.~\ref{fig:rq5c} reports the result of mAP~comparisons among these three kinds of models on the augmented~\textit{LISA} testing dataset (denoted with a `+' symbol, e.g., \texttt{RN+}) as well as on the cleaned augmented \textit{LISA} testing dataset (denoted with~a~`-'~symbol, e.g., \texttt{RN-}). First,~the original models suffer a smaller mAP decrease after removing unnatural data. This is reasonable as the original models are not trained with such unnatural data. Second,~on~the~augmented testing datasets, retrained models without removing unnatural data mostly have a slightly higher mAP than retrained models without unnatural data; i.e., the blue bars are mostly higher than the yellow bars. Hence, retrained models without removing unnatural data seems more robust. Third, on the cleaned augmented testing datasets, retrained models without removing unnatural data mostly have similar mAP as retrained models without unnatural data. This indicates that retraining models with those unnatural data will not affect models' capability on natural data.

Notice that here we only report the result of \textit{YOLOv5}.~The result of \textit{YOLOX} is similar, and is available at our website \url{https://tigaug.github.io/}.

\begin{tcolorbox}[size=title, opacityfill=0.15]
\textit{\textbf{Summary.}} \todo{27.9\%} of synthesized images by each transformation are considered~as~unnatural. However, these unnatural data, if used for retraining, will not affect models' capability on original real-world data and augmented natural data \ly{when all augmented training datasets are used together for retraining.}
\end{tcolorbox}

\subsection{Threats to Validity}

\textbf{Dataset Selection.} The quality and discrepancies of different driving~environments in the datasets pose a threat to validity. To that end, we select two well-known datasets, i.e., \textit{LISA} and \textit{Bosch}, which are widely used in the research community on autonomous driving systems~\cite{jensen2016vision, philipsen2015traffic, behrendt2017deep}. These datasets are collected in real-world~environments with varying light and weather conditions, and consist~of over 10k images and 20k annotated traffic lights. Therefore, both datasets cover a wide scope of driving environments. We believe~our evaluation results can be generalized to other datasets.

\textbf{Model Selection.} The performance discrepancies of~different~traffic light detection models affect the validity.~To~that~end, we select \textit{YOLOv5}, \textit{YOLOX}, \textit{Faster R-CNN} and \textit{SSD}~to~diversify~the~models~and our evaluation results. They are widely~used in the~research~community on autonomous driving~systems~\cite{pan2019traffic, kim2018deep, muller2018detecting, behrendt2017deep, li2021improved}. We obtain consistent~results from these four models, indicating that they can be generalized~to~other~models.

\textbf{Transformations.} We design and implement the 12 transformations under the consideration of generating traffic light images in road scenes as realistic as possible. In addition~to traditional considerations of weather conditions, we design~new transformations such as changing the color, position, and type of traffic lights with careful design. For example,~to~make~the generated images closer to real life, we only move traffic~lights by a small distance when changing the position of traffic~lights, ensuring that the traffic lights are still on the lamppost, rather than in other unreasonable places. By rotating the traffic lights, we transform the originally vertical traffic lights into horizontal traffic lights, which we believe has practical significance~in~real life.~Owing~to~our~careful design,~we~believe~our transformations are novel and reasonable. Besides, the~parameter~configuration in each transformation is empirically set to synthesize more natural images.~Therefore, our evaluation results might not generalize to other parameter configurations. We plan to continuously enlarge our transformation~set.

\textbf{Naturalness.} In computer vision literature, the ``naturalness'' of synthesized images is measured by some metrics~by~comparing the synthesized images and their corresponding original images, e.g., Structural Similarity Index Measure (SSIM)~\cite{Wang2004ImageQA}, Learned Perceptual Image Patch Similarity (LPIPS) \cite{Zhang2018TheUE}, and Histogram of Oriented Gradients (HOG) \cite{Dalal2005HistogramsOO}. However,~these metrics are used to measure the distance between the synthesized image and the original image, assuming that the more similar the synthesized image to the original image, the more natural the synthesized image is. This is not in line with~our~work as we may change the original image by a large distance which does not mean that we generate unnatural images.
%Structural Similarity Index Measure (SSIM)~\cite{Wang2004ImageQA} compares the structural, luminance and contrast information between the synthesized and original images to measure their perceptual similarity. Learned Perceptual Image Patch Similarity (LPIPS) \cite{Zhang2018TheUE} computes the perceptual similarity between the synthesized and original images by measuring the Euclidean distance between their deep feature embeddings. Histogram of oriented gradients (HOG) \cite{Dalal2005HistogramsOO} is a popular metric extracting distribution (histograms) of directions of gradients as “features” of an image. 
%Traffic light detection models are fed with real-world images captured by cameras. The unnaturalness of synthesized images threats the validity. To that end, 
\ly{During the transformation, we have taken the semantic relationships of traffic lights into account (i.e., in CC and RT). We also have certain constraints when changing positions of traffic lights (i.e., in MP and AD). However, there may still remain unnatural images.} Therefore, we design \textbf{RQ4}~to~manually exclude unnatural images and further investigate their impact. %As mentioned before, we remove the "unnatural" images through manual effort. Whether something is natural or not is quite subjective and is hard to judge, and it would require significant resources. 
We recruit three participants who are not informed of our applied transformations. We simply provide them with our synthesized images and ask them to decide whether they are natural or not. We find that~the large part of unnatural images synthesized~by~SW and~FG~are those that make traffic lights invisible. Even human~beings~cannot recognize the traffic lights, and it seems meaningless for models to detect. For LF, most of the synthesized unnatural~images are caused by adding lens flare to night images.~We~plan~to~improve LF to not augment night images. For CC, MP, AD and RT, the unnaturalness of their synthesized images is caused~by~the~oversized ground truth bounding box, which causes the fusion step~to~produce unnaturalness. Besides, we find that the unnatural images~seem~not affect models' detection capability on natural data. Therefore, we believe \tool's augmentation capability is useful. It is interesting to investigate the correlation between unnatural data and bad data.

%\textbf{Usage Scenario.} Our augmentation relies on the prior-knowledge of labeled traffic lights. Therefore, it cannot be used to augment unlabeled traffic light images. One potential remedy is to apply traffic light detection~models~before~augmentation.

% This is potentially because only 20\% of augmented training data is used, which is a common practice.

% !TeX root = ../main.tex

\section{Related Work}

%We review the related work in two areas: testing~autonomous driving systems and metamorphic testing for deep~learning.

%\subsection{Testing Autonomous Driving Systems}

\textbf{Testing Autonomous Driving Systems}. Testing has been recognized as one of the challenges in software~engineering for autonomous driving systems~\cite{Knauss2017, Czarnecki2019, Tahir2020}. Search-based testing has been widely investigated to find safety violations for autonomous driving systems~\cite{ben2016testing, abdessalem2018testinga, abdessalem2018testingb, gladisch2019experience, han2020metamorphic, li2020av, tian2022mosat, sun2022lawbreaker, gambi2019automatically, zhong2022detecting}.~They~often define a test scenario as a vector of multiple variables (e.g.,  vehicle speed), %, pedestrian position, and fog degree)
 and apply generic~algorithms to search the input space for test~scenarios that potentially violate a set of safety requirements. %(e.g., the distance between~the~vehicle and the pedestrian should be larger than a threshold). 
Differently, Tian et al.~\cite{tian2022generating} generate safety-critical test scenarios~by~influential behavior patterns mined from real traffic trajectories. 

Moreover, some advances have been made in testing DNN-based modules in autonomous driving systems. Pei et al.~\cite{pei2017deepxplore} design~a~neuron coverage-guided method, DeepXplore,~to~generate images~of~driving scenes by changing lighting conditions and occlusion with small rectangles. Tian et al.~\cite{tian2018deeptest} also~design a neuron coverage-guided approach, DeepTest, to generate images~of~driving scenes~more systematically through a more complete set of transformations. %(i.e., changing brightness, changing contrast, translation, scaling, horizontal shearing, rotation, blurring, fog effect, and rain effect).
 Different from them, Zhang et al.~\cite{zhang2018deeproad} use generative adversarial network~\cite{goodfellow2020generative} to synthesize images of driving scenes with various weather conditions. To work~with~dynamically changing driving conditions, Zhou~et al.~\cite{zhou2020deepbillboard} design DeepBillboard to generate adversarial perturbations that can be patched on billboards to mislead steering models. To test object detection models, Wang and Su~\cite{wang2020metamorphic} and Shao \cite{shao2021testing} propose to generate images~of~driving scenes by inserting extra object into background images.~To~test 3D object detection models, Guo et al.~\cite{guo2022lirtest} apply affine and weather transformation to augment LiDAR point clouds.

Besides, Gambi et al.~\cite{gambi2019generating} combine natural language~processing with a domain-specific ontology to automatically~generate car crash scenarios from police reports. Secci and Ceccarelli \cite{secci2020failures} summarize potential failure modes of a vehicle camera, and generate the corresponding failed images to analyze their effects on autonomous~driving systems. Deng et al.~\cite{deng2022scenario} slice a driving recording into segments, reduce segment length by removing redundancy, and prioritize the resulting segments based on coverage of driving scene features.~Despite these recent advances, many testing challenges still remain to better assure~safety of autonomous driving systems~\cite{tang2022survey, koopman2016challenges, lou2022testing}. 

\todo{To the best of our knowledge, except for Bai et al.'s work \cite{bai2021metamorphic},~none of the previous work is focused on the testing traffic light detection models. %, which are very critical DNN models in autonomous~driving systems. 
In Bai et al.'s work, the color of the traffic light is changed by labeling the traffic light bulb and assigning a single color (such as red (255, 0, 0))~to~the~bulb using certain rules. On the~contrary,~we~use HSV to change the color tone of traffic lights, which can make the generated images closer to real world and does not~require additional labeling of traffic light bulbs. 
Besides, while Bai~et~al. generate traffic light images~by~changing the color of traffic lights, our work introduces a more complete set of transformations~to~augment~traffic~light~images.}

%\subsection{Metamorphic Testing for Deep Learning}

\textbf{Metamorphic Testing for Deep Learning}. As metamorphic testing~\cite{chen2018metamorphic, segura2016survey} can alleviate the test oracle problem,~it~has~been widely used to test various deep learning~systems apart from autonomous driving systems. For example,~Dwarakanath et al.~\cite{dwarakanath2018identifying} use metamorphic testing to identify implementation bugs~in image classifiers. Sun et al.~\cite{sun2020automatic} combine mutation~testing with metamorphic testing to detect inconsistency bugs in machine translation systems. Chen et al.~\cite{chen2021validation} design~a~property-based method~to~validate machine reading comprehension~systems against seven metamorphic relations about different~linguistic properties. Liu et al. \cite{liu2021dialtest} use transformation-specific metamorphic relations with Gini impurity guidance to test dialogue systems. Ji et al.~\cite{ji2022asrtest} introduce three transformation families, i.e., characteristics mutation, noises injection, and reverberation simulation, based on metamorphic relations to test speech recognition systems. Yu et al.~\cite{yu2022automated} propose a metamorphic testing~approach~to validate image captioning~systems. %\todo{Our work shares the same nature of metamorphic testing, but targets a different domain.}

% !TeX root = ../main.tex

\section{Conclusions}
%The traffic light detection serves as a critical part in autonomous driving system (ADS). Current traffic light detection models heavily rely on the manually collected and labeled traffic light data, which is labor-intensive and time-consuming or even impossible to collect diverse data under different driving~environments. As a result, it hurts the effectiveness of traffic light detection testing. 

We have proposed a prototype tool,~named \tool, to automatically augment traffic light images for detecting erroneous behaviors and improving performance~of~traffic light detection models in ADSs. We have conducted~large-scale experiments~to~demonstrate~the effectiveness and efficiency of \tool. The source code of our tool and~our~experimental data are available at \url{https://zenodo.org/records/8213894}.

%%
% Example citation, See \cite{lamport94}.





% %% For numbered reference style

%   \textit{\LaTeX: a document preparation system},
%   Addison Wesley, Massachusetts,

% \end{thebibliography}

\bibliographystyle{model1-num-names}
\bibliography{src/reference}

\end{document}